\documentclass[10pt,aps,twocolumn,pre,floatfix,superscriptaddress]{revtex4-1}
\usepackage{graphicx}
\usepackage{epstopdf}
\usepackage{amssymb,amsfonts,amsmath}
\usepackage{color}
\usepackage{float}
\newcommand{\pd}[2]{\frac{\partial #1}{\partial #2}}
\newcommand{\mycom}[2]{\genfrac{}{}{0pt}{}{#1}{#2}}
\newcommand{\QualModel}{%
\begin{figure}[htbp]
        \begin{center}
                \includegraphics{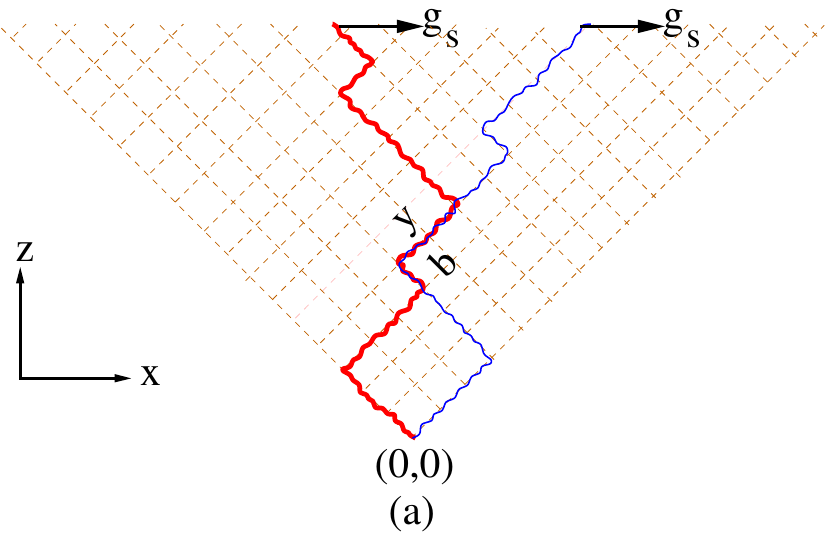}
                \label{fig0a}
                
                \includegraphics{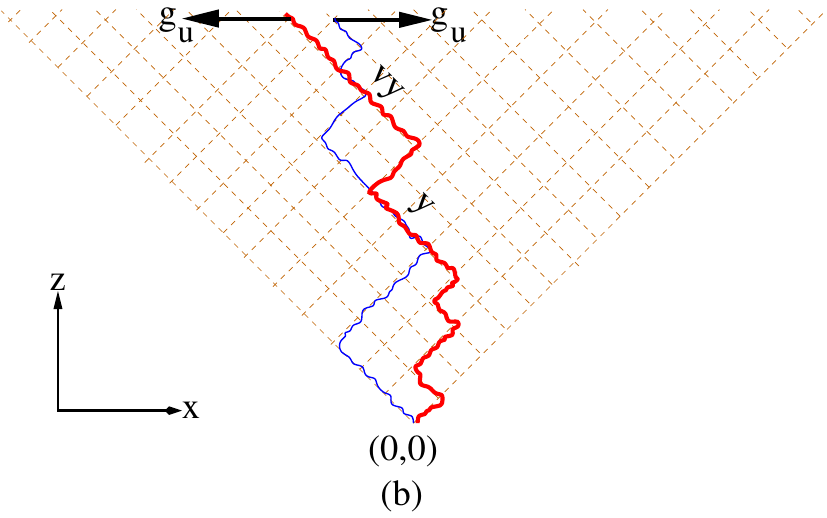}
                \label{fig0b}
        \end{center}
        \caption{(color online). Schematic of dsDNA
          as two directed walks in the $(1+1)$ dimension. The direction in
          which the forces act are indicated by arrows. (a)
          Flexible model : The polymers can not cross each other. The
          bound segments can bend left or right freely. This freedom
          can be restricted by introducing a statistical weight $b$.
          Here, $b$ is associated with the left degree of
          freedom. (b) Rigid model : The polymers can cross here. The
          bound segments can not bend to the right and they are at
          least two bonds long. $v$ is the weight associated with the
          bubble opening or closure.}\label{fig0}
\end{figure} }
\newcommand{\stfnc}{%
\begin{figure}[htbp]
  \begin{center}
   \includegraphics{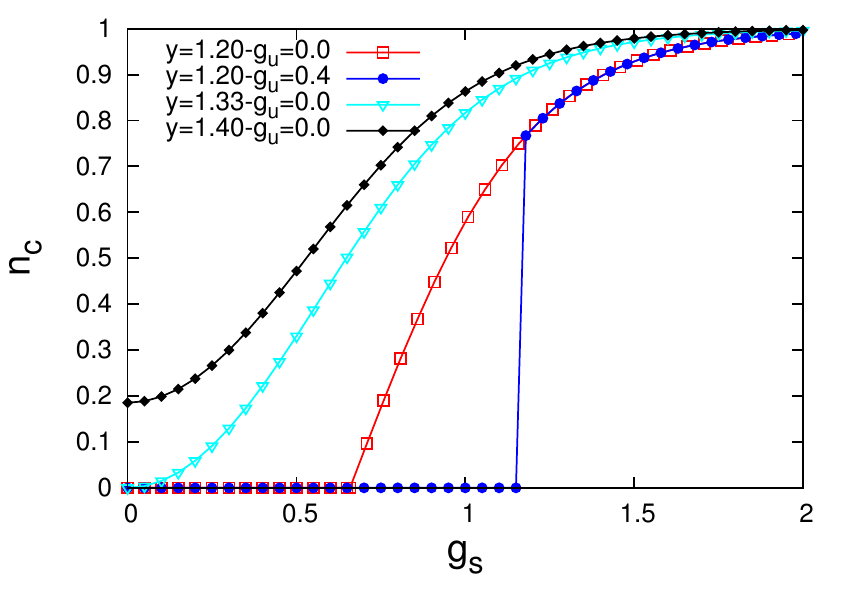}
 \end{center}
 \caption{ (color online). Flexible model : Variation of $n_c$
   with $g_s$ for different values of $g_u$ and $y$. A non-zero $n_c$
   indicates a bound state. For $g_u=0$, there is no unbound state for
   $y\geq4/3$. The $g_u=0$ curves are from Eqs. (\ref{EQ:5}),
   (\ref{EQ:11}), and (\ref{eq:11b}).  The cyan line with triangles is for
   the melting point $y=y_c$, and shows the quadratic dependence on
   $g_s$.  The black curve with filled diamonds represents $y=1.4>y_c$.
   The red curve, for $g_u=0$, shows a continuous transition, while the
   blue curve, for nonzero $g_u$, Eq. (\ref{EQ:14}), shows a
   discontinuity.  The symbols on these analytically obtained curves,
   in the infinite-chain limit, are to make them
   distinct.}\label{fig1}
\end{figure} }
\newcommand{\stfphdiag}{%
\begin{figure}[htbp]
  \begin{center}
   \includegraphics{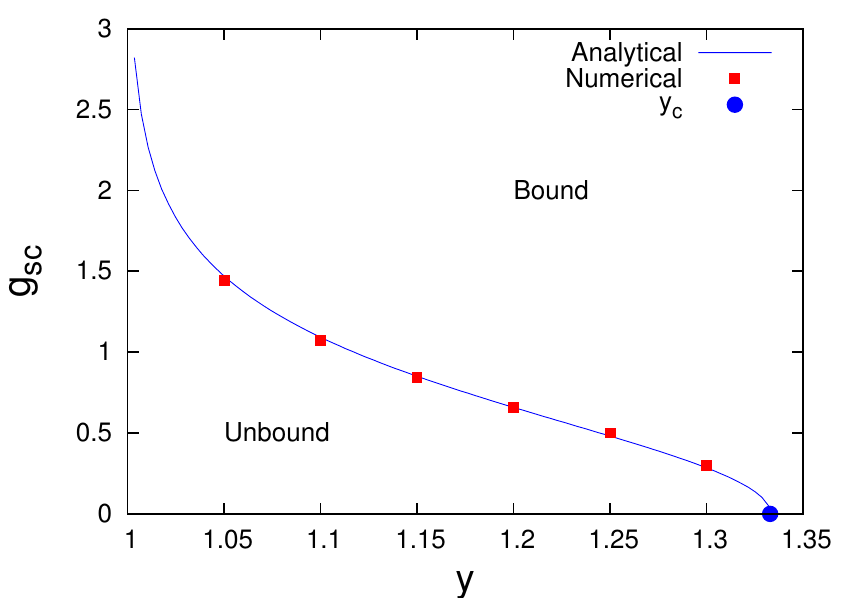}
 \end{center}
 \caption{(color online). Flexible model : Phase diagram of the
   system under a stretching force.  The phase boundary separates the
   bound phase from the unbound phase. In the region $4/3>y>1$ there
   exists a critical $g_{sc}$ for every value of $y$ . Above $y=4/3$
   the system is by default in the bound state. The solid blue line is
   the analytical curve, Eq. (\ref{EQ:12}), and the red squares represent 
   the numerically obtained critical points [see discussion following Eq.
   (\ref{scaling})].}\label{fig2}
\end{figure} }
\newcommand{\stfext}{%
\begin{figure}[htbp]
  \begin{center}
   \includegraphics{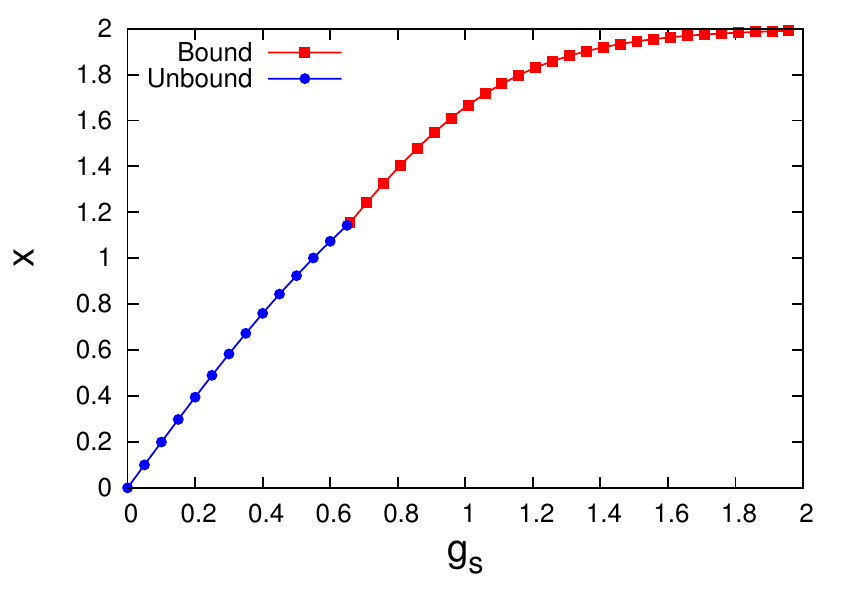}
 \end{center}
 \caption{(color online). Flexible model : Plot of the average
   extension as a function of the stretching force with a fixed
   $y=1.20$.  These are obtained from Eqs. (\ref{EQ:3}),
   (\ref{EQ:11}), and (\ref{eq:11b}).  The average extension varies
   continuously around the critical point.  The symbols on these
   analytically obtained curves are to make them
   distinct.}\label{fig3}
\end{figure} }
\newcommand{\figyzerog}{%
\begin{figure}[htbp]
  \begin{center}
   \includegraphics{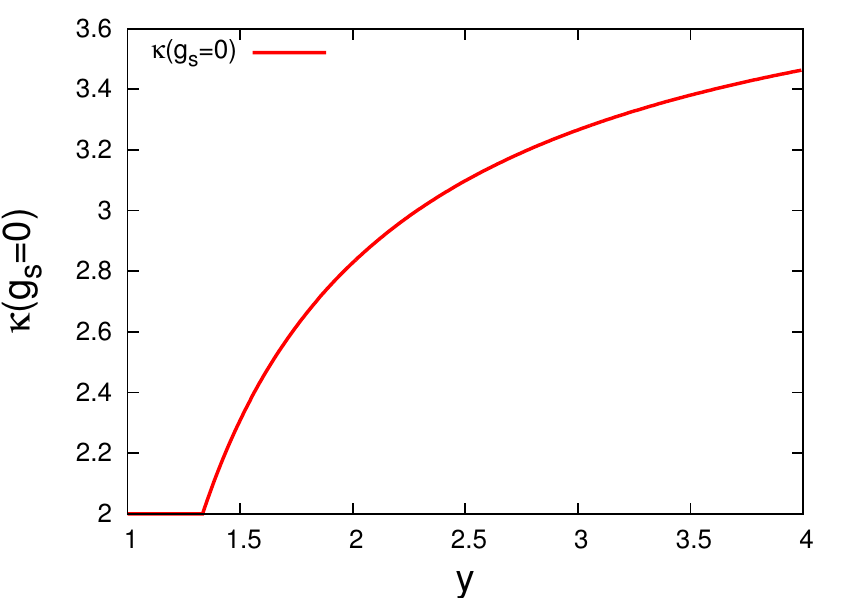}
 \end{center}
 \caption{(color online). Flexible model: The elastic modulus, Eq.
   (\ref{eq:1k}), as a function of $y$ for zero stretching
   force.}\label{figy1}
\end{figure} }
\newcommand{\figgsdy}{%
\begin{figure}[htbp]
  \begin{center}
   \includegraphics{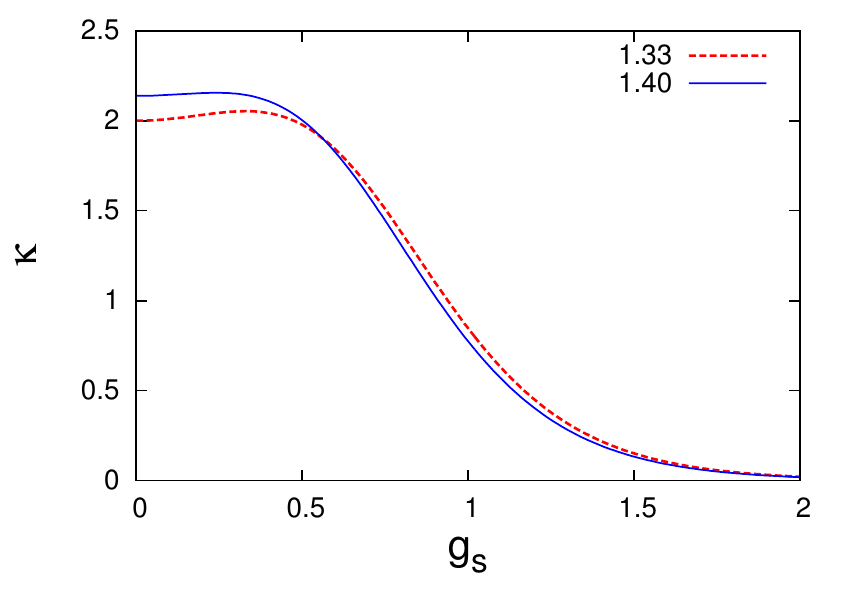}
 \end{center}
 \caption{(color online). Flexible model: The elastic modulus as a
   function of $g_s$ at $y=4/3=y_c$ [Eq. (\ref{eq:2c}] and
   $y=1.40>y_c$ [from Eqs. (\ref{EQ:10a}) and (\ref{EQ:10b})].}\label{figy2} 
\end{figure} }
\newcommand{\ananumcoma}{%
 \begin{figure}[htbp]
 \begin{center}
  \includegraphics{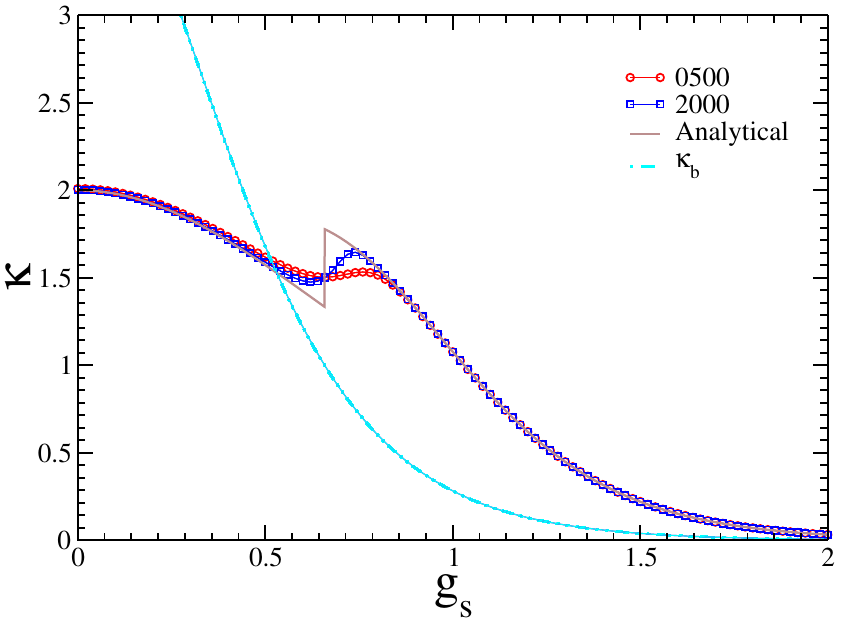}
  \end{center}
  \caption{(color online). Flexible model : Elastic modulus curve
    for $y=1.20$. The solid brown line is the analytically obtained
    curve (see text).  The dashed cyan line is for $\kappa_b=4\
    \text{sech}^2(2g_s)$, Eq.  \eqref{ymod1}, for the case with no
    bubbles.  Other curves are the plots of $\kappa$ for different
    system lengths.  Solid lines through the data points are
    guides for the eye.}\label{fig4a}
 \end{figure}}
 \newcommand{\ananumcomb}{%
 \begin{figure}[htbp]
 \begin{center}
  \includegraphics{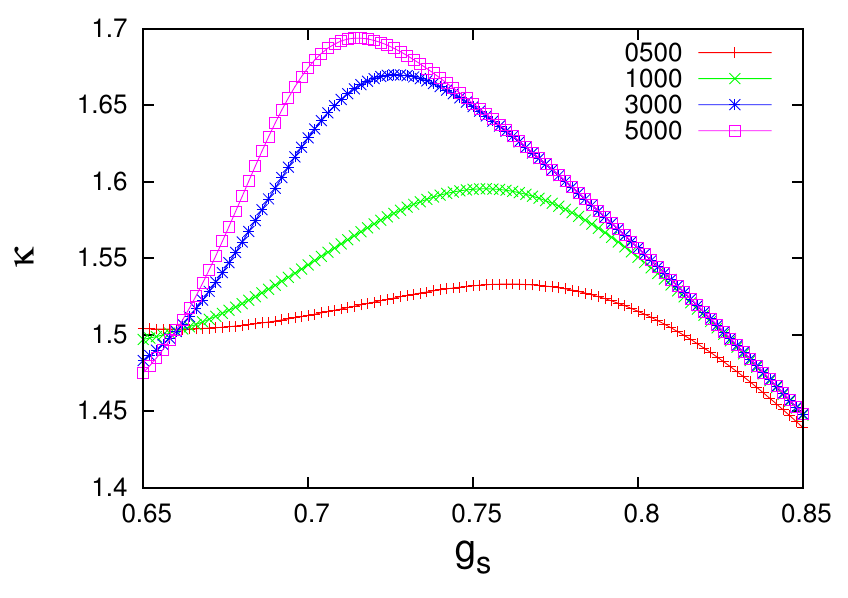}
  \end{center}
  \caption{(color online). Flexible model : A magnified version of
    Fig.  \ref{fig4a} around the same crossing point of all the
    curves.  The chain length for each curve is given in the legend.
    Solid lines through the data points are guides for the
    eye.}\label{fig4b}
\end{figure} }
\newcommand{\unzanaext}{%
 \begin{figure}[htbp]
 \begin{center}
     \includegraphics{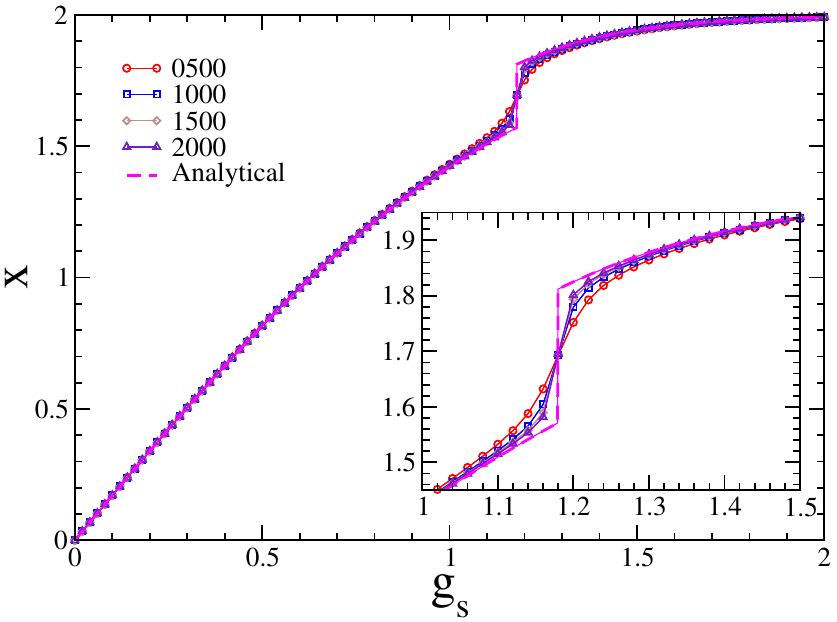}
 \end{center}
 \caption{(color online). Flexible model: $x$ vs $g_s$ plot, with
   $g_u=0.4$ and $y=1.20$. Inset: Magnification of the critical region.
   The dotted magenta line is the analytical curve. See Sec.
   \ref{sec:elastic-constant}.  All other curves are for finite
   system sizes shown in the legend. The discontinuity at $g_{sc}$
   indicates a first order transition.  Solid lines through the
   data points are guides for the eye.}\label{fig6a}
 \end{figure}}
 \newcommand{\unzanelc}{%
 \begin{figure}[htbp]
 \begin{center}
   \includegraphics{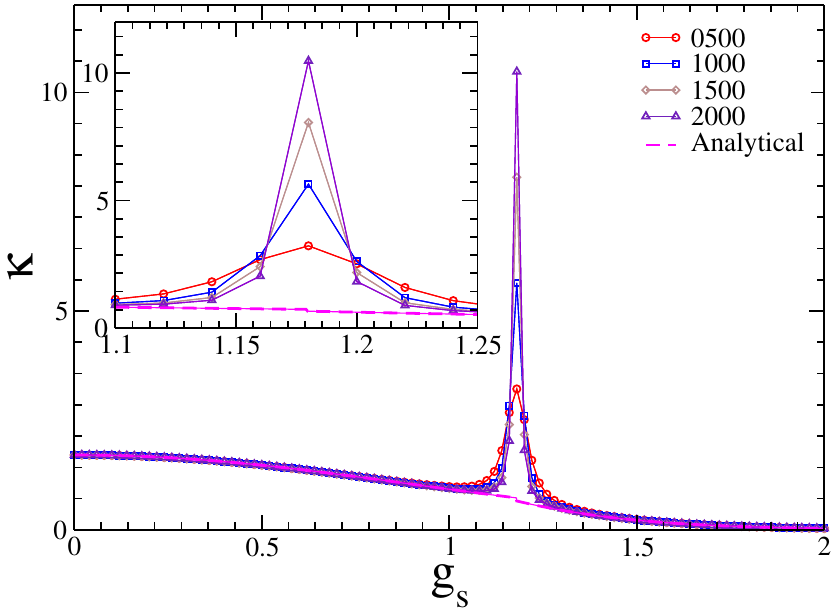}
 \end{center}
 \caption{(color online). Flexible model: $\kappa$ vs $g_s$ plot with
   $g_u=0.4$ and $y=1.20$. Inset: Magnification of the critical region.
   The dotted magenta lines are the analytical curves.  See Sec.
   \ref{sec:elastic-constant}.  All other curves are for finite
   system sizes shown in the legend.  The peak height increases
   proportionally with $N$, signaling a $\delta$-function peak which is
   not shown in the analytical curve.  Solid lines through the
   data points are guides for the eye.}\label{fig6b}
\end{figure}}
\newcommand{\threedphdiag}{%
\begin{figure}[htbp]
  \begin{center}
   \includegraphics{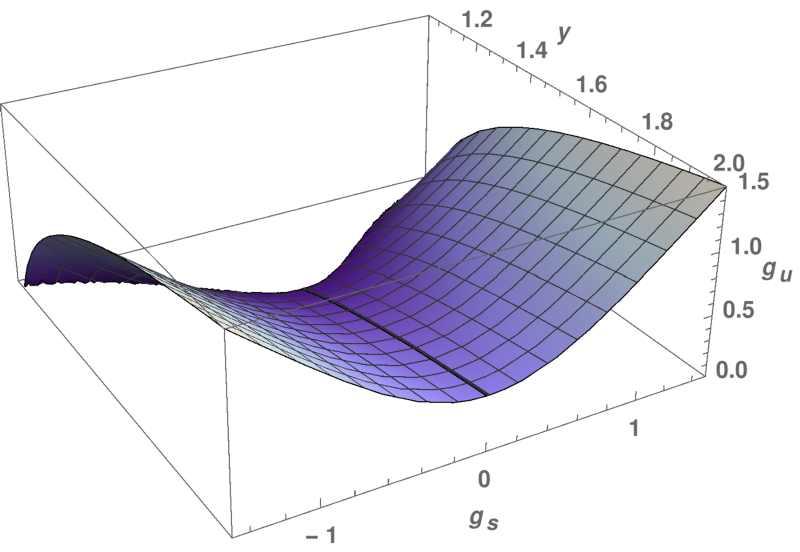}
 \end{center}
 \caption{(color online). Flexible model: Three dimensional phase
   diagram of the system, Eq. (\ref{EQ:15}), in the presence of both
   stretching and unzipping forces.  All points on the surface
   except the curve for $g_{uc}=0$ represent first-order phase
   transition points. The unzipping line for $g_s=0$ is shown by the
   thick black line.}\label{fig7}
\end{figure} }
\newcommand{\rcnc}{%
\begin{figure}[htbp]
  \begin{center}
   \includegraphics{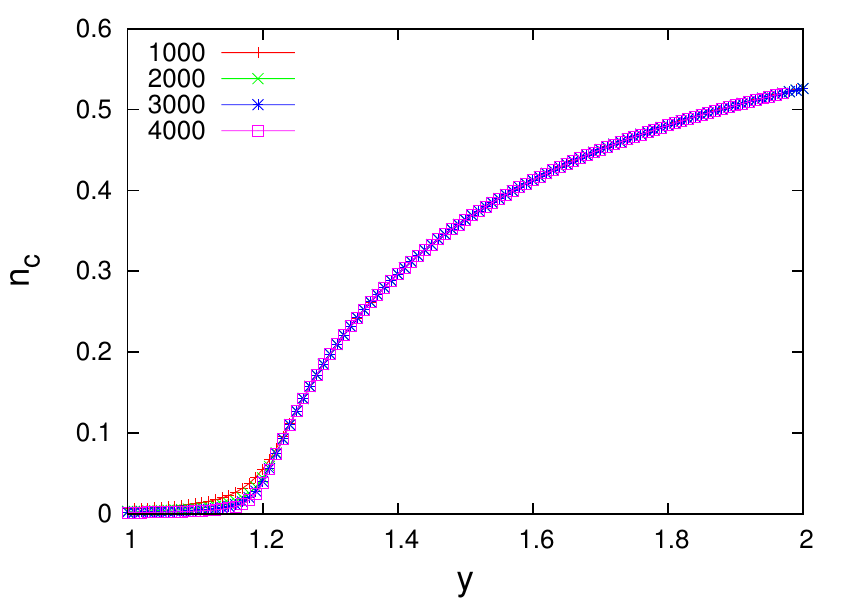}
 \end{center}
 \caption{(color online). Rigid model: $n_c$ increases from zero to
   non-zero values continuously at $y>1$ before approaching the
   saturation value, 1. This indicates a binding-unbinding transition.
   Solid lines through the data points are guides for the
   eye.}\label{fig8}
\end{figure} }
\newcommand{\rcsp}{%
\begin{figure}[htbp]
  \begin{center}
   \includegraphics{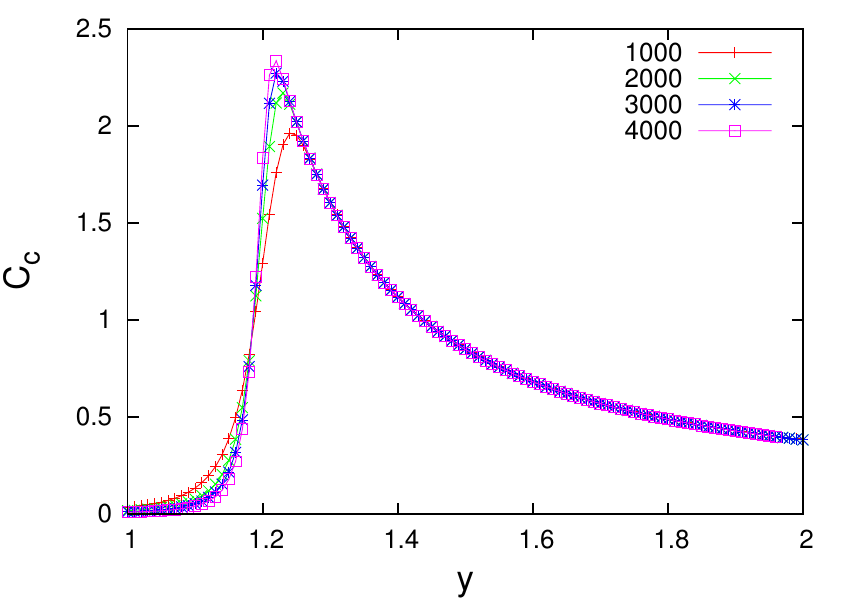}
 \end{center}
 \caption{(color online). Rigid model: $C_c$ vs $y$ plots for
   different $N$. The peak height increases with increasing $N$ but
   eventually saturates, creating a finite discontinuity.  The
   discontinuity occurs at $y=1.18$, where all the curves meet. 
   Solid lines through the data points are guides for the
   eye.}\label{fig9}
\end{figure} }
\newcommand{\rcext}{%
\begin{figure}[htbp]
  \begin{center}
   \includegraphics{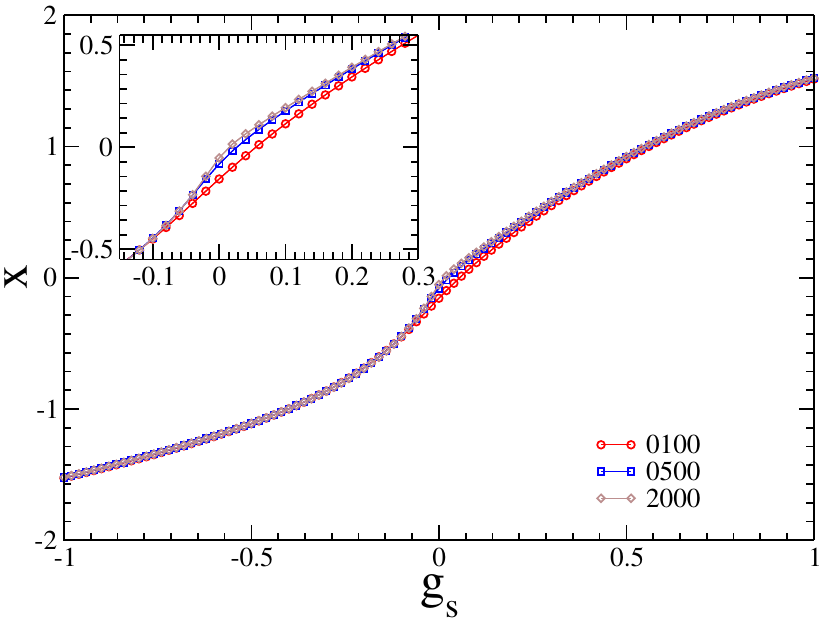}
 \end{center}
 \caption{(color online). Rigid model: Continuous stretching of the
   DNA to its full extent by both positive and negative forces. The $y$
   value is fixed at 1.20. Solid lines through the data points are
   guides for the eye.}\label{fig10}
\end{figure} }
\newcommand{\rcelc}{%
\begin{figure}[htbp]
  \begin{center}
   \includegraphics{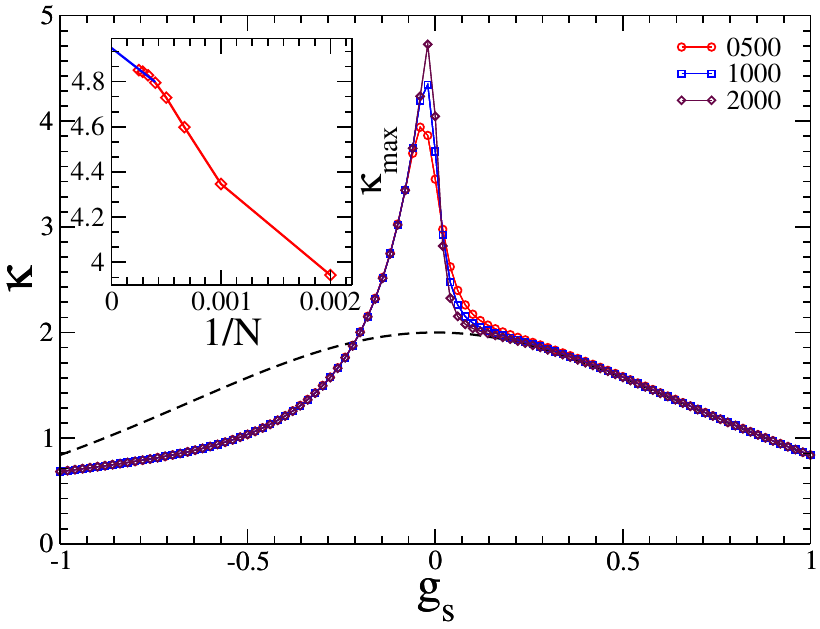}
 \end{center}
 \caption{(color online). Rigid model: $\kappa$ shows an increasing
   peak around $g_s=0.02$ with increasing system length $N$ and a
   fixed $y=1.20$. Inset: Maximum values of $\kappa$, $\kappa_{\text{max}}$,
   are plotted vs $1/N$. A linear fit with the fist four
   points (solid blue line) gives the estimate for $N\to\infty$,
   $\kappa_{\text{max}}=4.95$. This indicates that there is a finite
   discontinuity in $\kappa$. The dashed black line is a plot of the
   function $\kappa=2\ \text{sech}^2(g_s)$, the unbound-state elastic
   constant. For $g_s>0.02$, $\kappa$ matches the unbound-state
   modulus.  Solid lines through the data points are guides for the
   eye.}\label{fig11}
\end{figure} }
\newcommand{\rcfnc}{%
\begin{figure}[htbp]
  \begin{center}
   \includegraphics{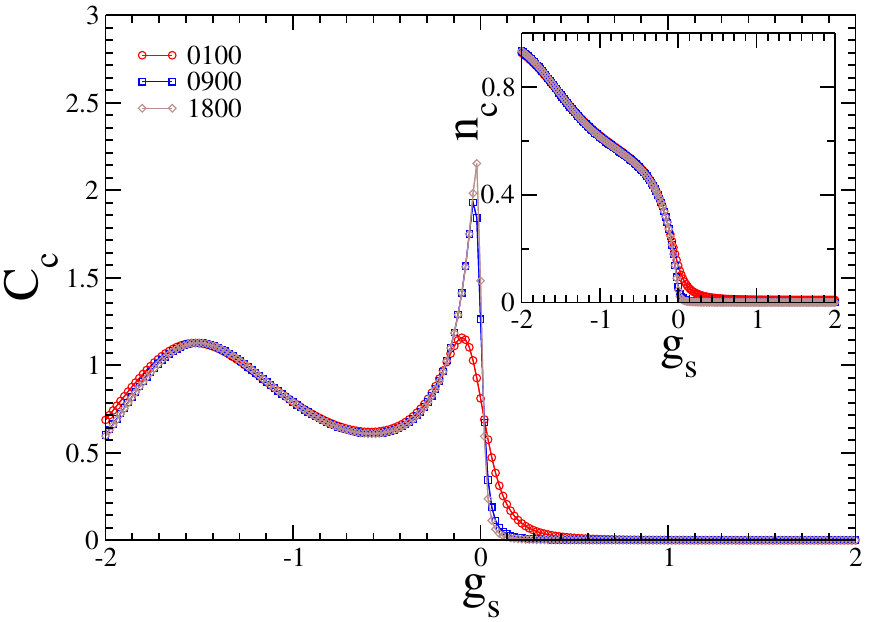}
 \end{center}
 \caption{(color online). Rigid model: $C_c$ vs $g_s$ curves for
   finite lengths as indicated. The $y$ value is fixed at 1.20. 
   Inset: $n_c$ decreases continuously from a finite value
   to 0, indicating a continuous phase transition. Peak heights in
   $C_c$ vs $g_s$ curves saturate, indicating a finite discontinuity.
   Around $g_s=0.02$ all curves meet at the critical point.  Solid
   lines through the data points are guides for the eye.}\label{fig12}
\end{figure} }
\newcommand{\rcfphd}{%
\begin{figure}[htbp]
  \begin{center}
   \includegraphics{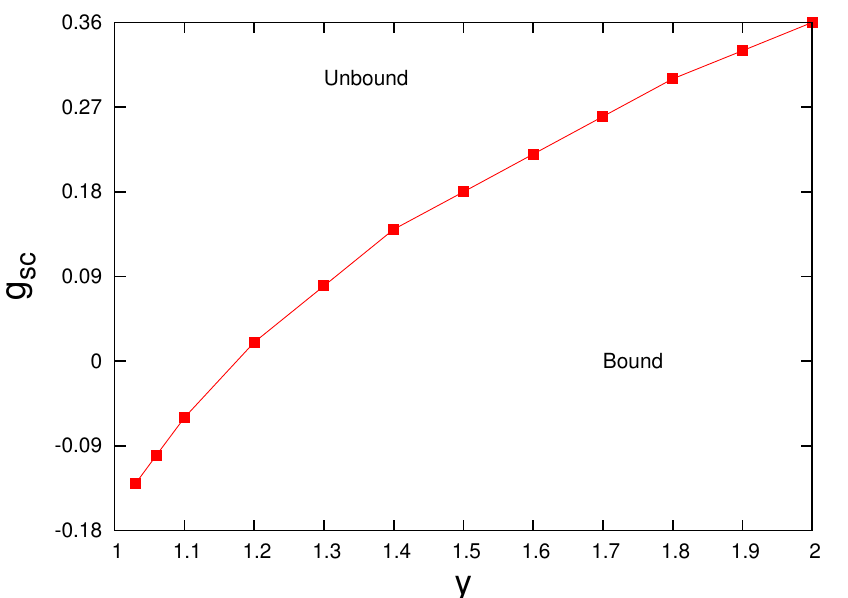}
 \end{center}
 \caption{(color online) Rigid model: Numerical phase diagram for the
   binding-unbinding transition.  The solid line through the data
   points is guide for the eye.}\label{fig14}
\end{figure} }
\newcommand{\rcfnblb}{%
\begin{figure}[htbp]
  \begin{center}
   \includegraphics{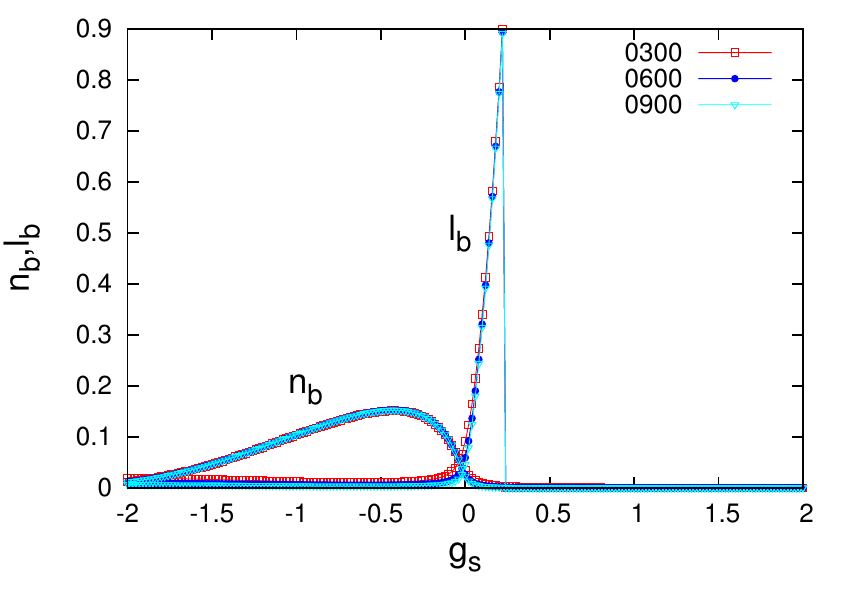}
 \end{center}
 \caption{(color online). Rigid model: $n_b$ vs $g_s$ and $l_b$ vs
   $g_s$ plot for $y=1.20$.  As the critical point is approached,
   $n_b$ becomes very small but $l_b$ becomes as large as $N$. 
   Solid lines through the data points are guides for the
   eye.}\label{fig15}
\end{figure} }
\newcommand{\rcfnbcb}{%
\begin{figure}[htbp]
  \begin{center}
   \includegraphics{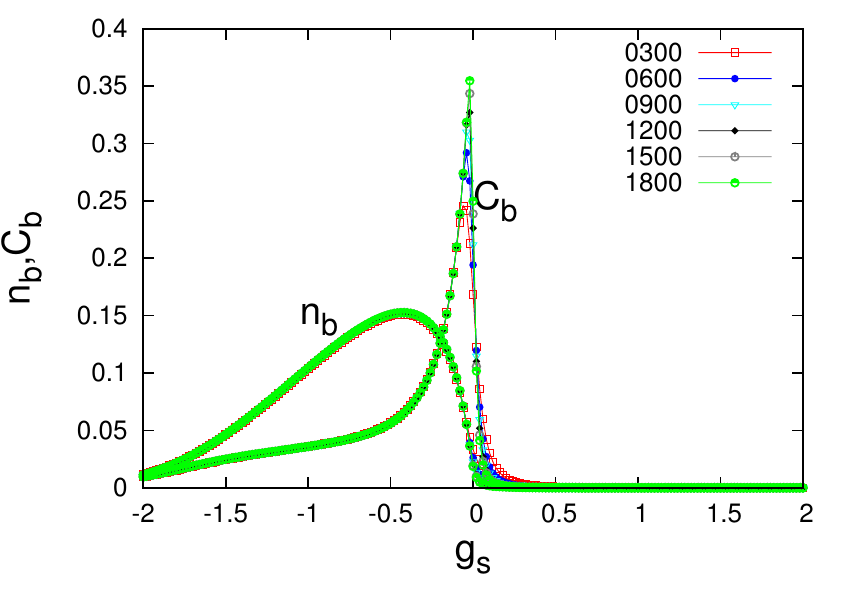}
 \end{center}
 \caption{(color online). Rigid model: $n_b$ vs $g_s$ and $C_b$ vs
   $g_s$ plot for $y=1.20$.  Around the critical point $n_b$ is very
   small but its fluctuation $C_b$ is very large.  Solid lines
   through the data points are guides for the eye.}\label{fig16}
\end{figure} }
\newcommand{\biasednc}{%
\begin{figure}[htbp]
  \begin{center}
   \includegraphics{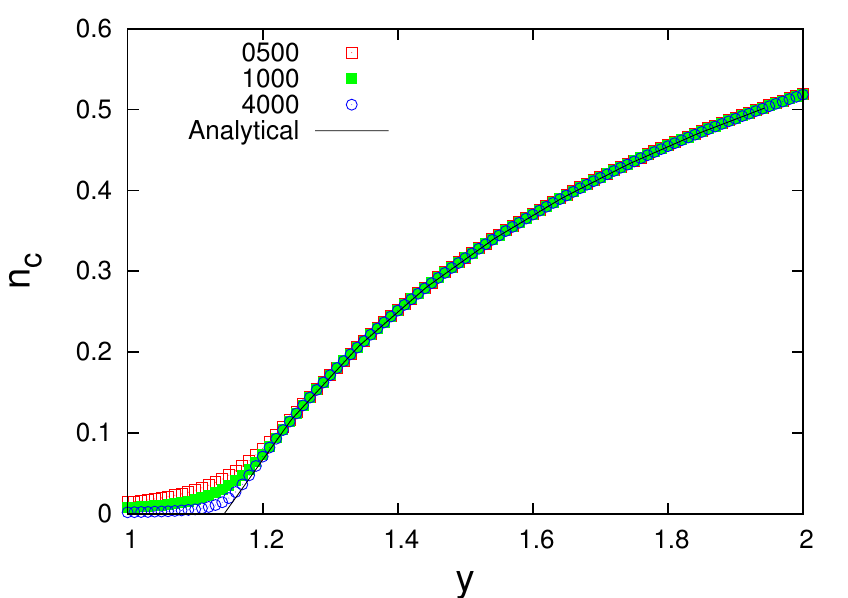}
 \end{center}
 \caption{(color online). $n_c$ vs $y$ plot for $b=0.5$. $n_c$ becomes
   finite at $y=1.142$.  Numerical data are also consistent with the
   analytical result.  Solid lines through the data points are
   guides are the eye.}\label{figx1}
\end{figure} }
\newcommand{\biasedsp}{%
\begin{figure}[htbp]
  \begin{center}
   \includegraphics{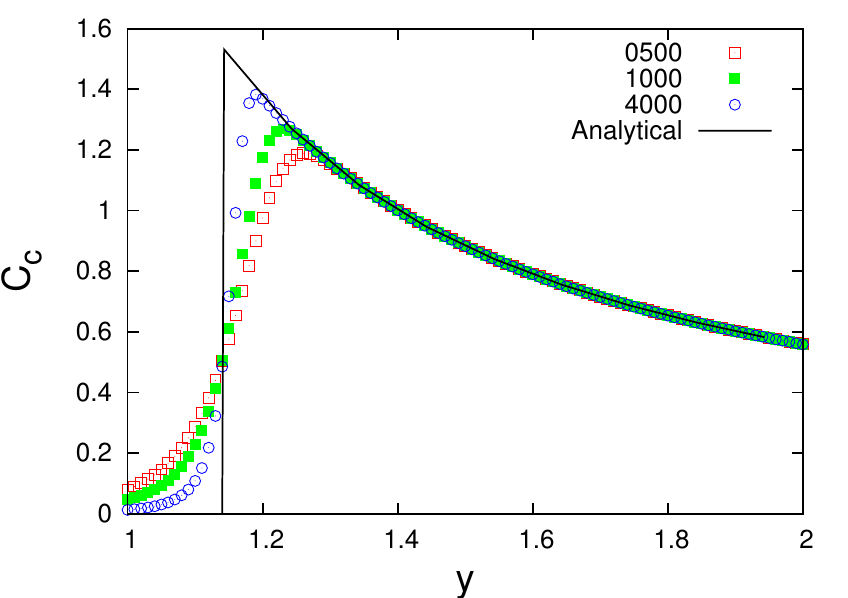}
 \end{center}
 \caption{(color online). $C_c$ vs $y$ plot for $b=0.5$ showing a
   finite discontinuity at $y=1.142$. Numerical data also show very
   similar behavior.  Solid lines through the data points are
   guides for the eye.}\label{figx2}
\end{figure} }
\newcommand{\xnr}{%
\begin{figure}[htbp]
  \begin{center}
   \includegraphics{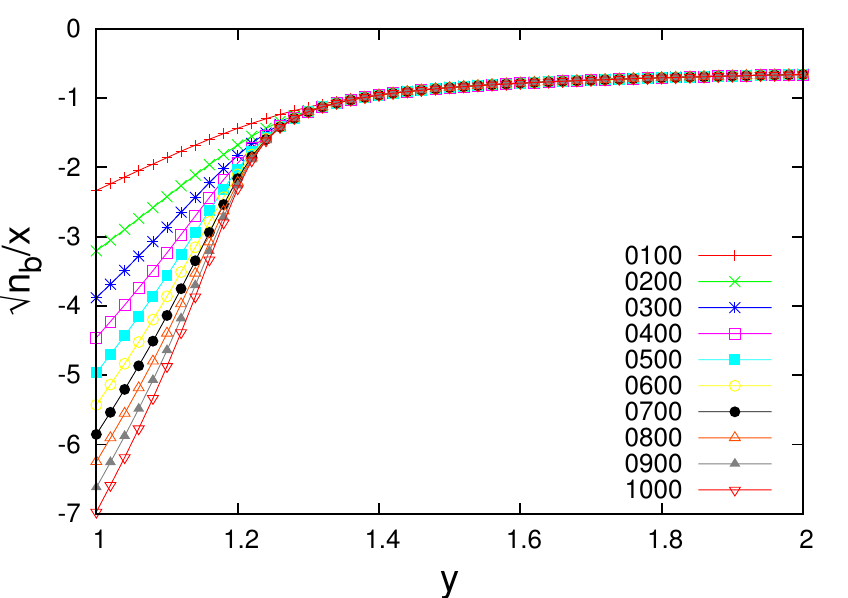}
 \end{center}
 \caption{(color online). Rigid model: $\frac{1}{x}\sqrt{n_b}$ vs $y$
   curves for different system lengths collapse to a single master
   curve in the bound region.  $g_u = g_s =0$. $y_c=1.18$.  Solid 
   lines through the data points are guides for the eye.}\label{fig17}
\end{figure} }
\newcommand{\kcr}{%
\begin{figure}[htbp]
  \begin{center}
   \includegraphics{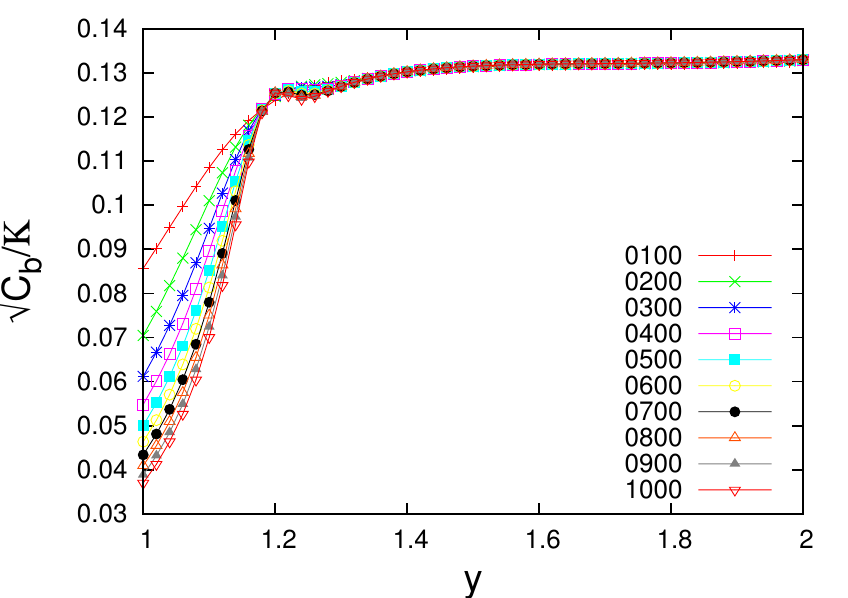}
 \end{center}
 \caption{(color online). Rigid model: $\frac{1}{\kappa}\sqrt{C_b}$ vs
   $y$ curves for different system lengths collapse to a single master
   curve in the bound region.  $g_u = g_s =0$. $y_c=1.18$.  Solid 
   lines through the data points are guides for the eye.}\label{fig18}
\end{figure} }
\begin{document}
\title{Rigidity of Melting DNA}

\author{Tanmoy Pal}
\affiliation{Institute of Physics, Bhubaneswar 751005 India}
\email[email: ]{tanmoyp\string@iopb.res.in} 
\author{Somendra M. Bhattacharjee} 
\affiliation{Department of Physics,
  Ramakrishna Mission Vivekananda University, PO Belur Math, Dist.
  Howrah, West Bengal 711202, India} 
\email[email: ]{somen\string@iopb.res.in; somen\string@rkmvu.ac.in}
\affiliation{Institute of Physics, Bhubaneswar 751005 India}

\begin{abstract}
The temperature dependence of DNA flexibility is studied in the presence 
of stretching and unzipping forces. Two classes of models are 
considered. In one case the origin of elasticity is entropic due to the 
polymeric correlations, and in the other the double-stranded DNA is taken to 
have an intrinsic rigidity for bending. In both cases single strands are 
completely flexible. The change in the elastic constant for the flexible 
case due to thermally generated bubbles is obtained exactly. 
For the case of intrinsic rigidity, the elastic constant is found to be 
proportional to the square root of the bubble number fluctuation. 
\end{abstract}

\maketitle

\section{Introduction}
To facilitate different fundamental biological processes, like
replication, gene expression, assembly of functional nucleoprotein
structures, and packaging of viral DNA, DNA has to go through a
lot of twisting, stretching, and bending
\cite{bend-1,bend-2,giese,schultz,nagaich,ortega,watson}. Generally
different proteins induce these conformational changes in DNA, but not
without facing any resistance. This is because, when subjected to an
external mechanical force, DNA responds elastically.  Single-stranded
DNA may be flexible and easy to bend, but double-stranded DNA (dsDNA)
is known to be more rigid. However, the flexibility of short DNA
fragments is important for different {\it in vivo} mechanisms, like
those already mentioned, where loops or bends as short as
100 base pairs in length are involved \cite{bend-3,bend-4}, and also in {\it in
  vitro} experiments, where fragments are used in open or hairpin
geometries. It is therefore important to probe the elastic response of 
dsDNA not only in the thermodynamic limit of long length --- dsDNA
is long --- but also for finite-size systems.

Topological arguments, {\it a la} the C{\u{a}}lug{\u{a}}reneau theorem
\cite{volod}, indicate the necessity of two major elastic constants of 
dsDNA, namely, the twist and the bending elastic constants. The
former is tied to the helical nature of the double-helix and the
latter is determined by both entropy and angular interactions between
neighboring tangent vectors. These elastic moduli are characteristics
of the bound phase and they disappear on DNAs melting into the
denatured phase. It is quite analogous to the disappearance of the
shear elastic modulus of a crystalline solid upon melting into the
liquid phase. If dsDNA is treated as a free Gaussian polymer with
noninteracting monomers, even then it exhibits an entropic elasticity
originating from the correlations of a random walk. On the other hand,
an intrinsic rigidity against bending at short scales (a semi flexible
chain) produces a temperature dependent persistence length ($\sim150$
base pairs), within which a dsDNA acts more or less like a rigid rod.
Thus, it seems, a larger force is required to bend DNA of a length
shorter than its persistence length than DNA of a longer length
\cite{doi,smb-flory}.  Recent debates
\cite{cloutier,wiggins,yanmarko,menon,menon1,qdu,mazur,noy,linna,shroff,forties,le,kramenetskii}
on the behavior of short segments brought into focus the importance of
broken base pairs on its eventual or effective rigidity. As the base-pair 
energy is comparable to the thermal energy at physiological
temperatures ($\sim$2-3 kcal/mol), bubbles form spontaneously or are
produced by external forces (see, e.g., \cite{bubble-force} for an
earlier study).  Consequently, the issue of the elastic response of 
dsDNA cannot be studied in isolation as its intrinsic property but
rather needs to be coupled to the inner degrees of freedom, namely, 
base pairings responsible for the bound state.

Breaking the base pairings can separate dsDNA into two
independent single strands and this melting happens at a particular
temperature. The thermal melting of DNA is by itself an interesting
problem and important for different {\it in vivo} or {\it in vitro}
processes. A notable example is the polymer chain reaction, which
is used extensively in DNA amplification. Other than that it has been
proposed \cite{efimov1,efimov2,efimov3} that at the dsDNA melting
point the addition of a third strand may support a three-stranded DNA
bound state where no two pairs of strands are expected to be bound.
This novel three-stranded DNA bound state is called Efimov-DNA and has been 
shown to support a renormalization-group limit cycle \cite{tp1,tp2}.
In the temperature region below the melting point there can be local
melting at different positions, creating denaturation bubbles, which are
nothing but single stranded loops preceded and succeeded by double-stranded 
segments. As single-stranded DNA is far more flexible than paired ones, 
these thermally generated bubbles can provide flexible
hinges which can make dsDNA significantly flexible
\cite{yanmarko,menon,menon1}. Generally the average length of these
bubbles increases as the critical point is approached and equals the
system length at the melting temperature.  The importance of bubbles
for the melting transition is well understood in the Poland-Scheraga
framework \cite{poland}. The entropic contributions of the bubbles in
different models, originating from long range polymeric
correlations of the individual strands, lead to different types of
melting behavior, from weakly first order to infinite order
\cite{peyrard1,peyrard2,peyrard3,prakash,efimov1,efimov2,efimov3,tp1,tp2,poland,kafri1,
  kafri2,trovato,marenduzzo,azbel}.  The simpler coarse-gain level
models show critical behavior
\cite{poland,peyrard1,peyrard2,peyrard3,trovato,fisher,tp1,tp2,sutapa,poulomi}.
Close to melting, the bubbles are therefore expected to contribute
significantly to the flexibility, beyond just acting as hinges.
Different from thermal melting is the unzipping transition, where the
two strands of dsDNA are pulled apart at temperatures below the
melting point. The unzipping transition is generally first order
\cite{smb-unzip,poulomi}, even in models with a first order melting
transition \cite{kafri1,kafri2}.  Since the unzipping force does not
penetrate the bound state \cite{poulomi,skumar}, the nature of the
bubble distribution does not change in the presence of an unzipping
force. As a result, the bubbles are going to have their signatures on
the flexibility of a DNA near the unzipping transition too. From a
phase transition point of view, the bending elastic constant, despite
its importance for DNA activity, is not a primary response function
that characterizes a critical point. For example, one may compare it with the
magnetic susceptibility of a ferro-para magnetic transition or the
elastic modulus in a liquid-gas transition, whose divergence is
associated with an exponent $\gamma$. But as it is not the primary response function, 
no such general results of critical phenomena are applicable here.
Hence the necessity for a detailed study of the rigidity of melting
DNA. In this paper we want to explore the elastic properties of DNA
near its melting point. DNA melting is a genuine phase transition
for which the DNA length has to satisfy the thermodynamic limit. But
still, the existence of a transition point is sufficient to affect a
finite-size system even when it is away from the critical point. One
of our aims is to obtain a few exact results on the elastic behavior
for a class of models of DNA melting and unzipping.

Different statistical mechanical models have been applied with varying success to 
study the DNA elasticity problem. The classical semiflexible chain model with 
no denaturation bubbles has been employed by a number of investigators 
\cite{zimm,seol}. Segments made of single strands can be introduced in 
discretized semiflexible DNA, by considering models comprised of two-state 
internal coordinates and, also, by coupling these internal coordinates to the 
external rotational degrees of freedom of its tangent 
vectors \cite{nikos,kwlc,palmeri}. Bubbles appear naturally in our models 
without utilizing any other secondary variables. 

The modulus of interest comes from the response to a force applied at one end 
of each of the two strands, keeping the other end fixed at the origin. 
We use models of DNA where the strands are represented as polymers 
with native base pairing; namely, two monomers of the two strands interact 
only when they have the same contour length on the polymer. To probe the 
elastic behavior, we use a stretching force that would act on both 
strands in the same direction, while the phase can be changed by an 
unzipping force that acts on the strands in opposite directions. Here 
we quantitatively relate DNA flexibility to bubble-related quantities 
such as the bubble length, the bubble number etc.

The organization of this paper is as follows. 
In Sec. \ref{secII}, we qualitatively describe the models, 
the flexible model and the rigid model, considered in this paper. 
Section \ref{secIII} is devoted to the flexible model. In Sec. 
\ref{secIIIA}, we introduce the corresponding recursion relations and 
define the observables necessary for analysis of both 
models. The elastic properties of the flexible model are explored 
in Sec. \ref{secIIIB}, where we show a finite discontinuity in the elastic 
modulus at the melting point. We obtain the phase diagram in the presence of an 
unzipping force and a stretching force in Sec. 
\ref{secIIIC}. The transition is now first order and the elastic modulus 
shows a $\delta$-function peak at the transition point. We introduce the 
rigid model in Sec. \ref{secIV}. After listing its 
governing recursion relations and defining the required observables specific 
to this model, we obtain its thermal melting (a continuous transition) 
point in Sec. \ref{secIVA}. In Sec. \ref{secIVB}, we show that the 
corresponding elastic modulus becomes anomalous around the melting 
point as it surpasses the unbound-state elastic modulus. The roles played 
by the bubbles are shown quantitatively in Sec. \ref{secIVC}. Only the 
stretching force is considered for the rigid model. After a brief 
discussion of the relevance of our results in Sec. \ref{secV}, we 
summarize and conclude in Sec. \ref{secVI}. A few important supplementary 
materials are listed in the Appendices \ref{appnA} and \ref{appnB}. 
\section{Qualitative Description}\label{secII}
\QualModel
To isolate the entropic and the intrinsic elastic behavior, two 
types of models are considered, viz., a flexible model,
the standard model used for melting and unzipping
\cite{marenduzzo,rajeev12,rajeev11,trovato,rajeev}, and a rigid model, built
from the flexible model. See Fig. \ref{fig0}. Both models show a 
zero-force melting point, generically denoted $y_c$.
The common features of these models are as follows : we consider 
each single strand as a directed polymer in a two-dimensional 
square lattice. We represent the base pairing by contact interactions 
between the monomers which can only occur at the same space and length 
coordinates. Correct base-pair bonding is ensured by the directedness of 
the polymers. The chains are inextensible \cite{sdna}, of the same length, and attached to 
each other at the origin. We mimic DNA melting by the 
binding-unbinding phase transition in the system. The statistical 
weight of a contact interaction is the Boltzmann factor 
$y=\exp(\beta\epsilon)$, $-\epsilon$ being the energy per contact and 
$\beta$ the inverse temperature with the Boltzmann constant set to $k_B=1$. 
Such models in the past have been instrumental in studies of the melting and 
the unzipping transition of dsDNA and are known to show the relevant features 
of the higher dimensional models \cite{smb-unzip,trovato,rajeev11,rajeev12}. 
These models are also useful for studies of the dynamics of DNA. The flexible model 
[Fig. \ref{fig0}(a)] has a hard-core repulsion that 
forbids the two chains to cross. The perfectly bound DNA remains as flexible 
as the single-stranded DNA so that at any nonzero temperature there is only the 
emergent entropic elasticity. In contrast, the rigid model [Fig. \ref{fig0}(b)] 
has a bound state which has an intrinsic rigidity towards or against bending. 
In fact we consider the dsDNA to be absolutely rigid in the bound state. The 
only way it can show any flexibility is via denaturation bubbles. Thus, 
the elastic response in this model is purely due to the flexible hinges made 
accessible by the bubbles. It is possible to allow some controlled 
semiflexibility, instead of absolute rigidity, via the 
introduction of a parameter $b$, which penalizes the bound DNA taking an 
unfavorable turn. This model is discussed the Appendix \ref{appnB}. This 
imposed rigidity is enough to give a melting transition even in the absence of 
any hard-core constraint. To incorporate this rigidity unambiguously, 
a constraint is required that a bound base pair can form only if its 
previous monomers are in the bound state.

We apply an external space-independent mechanical 
stretching force independently at the end of each strand. If the two forces are 
in the same direction, the dsDNA is said to be under a stretching force $g_s$. 
On the other hand, it will be under an unzipping force $g_u$ if the forces are in 
opposite directions. The average extension and the elastic modulus can be obtained 
from the free energy simply by taking derivatives with respect to the stretching 
force once and twice, respectively.

We use the transfer matrix method through recursion relations to find the partition 
function of the system. For the analytical solution, the generating function 
for the grand partition function is used, from whose 
singularities the free energy can be determined \cite{marenduzzo}. For numerical 
calculations we iterate the recursion relations for finite lengths and find the 
exact partition function. The numerical calculations reflect the finite-size 
behavior of the concerning quantities. The effect 
of the unzipping force in the elasticity is also explored. The general case 
of two unequal forces can always be transformed into a case of unzipping and 
stretching forces. Since unzipping and stretching of dsDNA are independent 
of each other, we are able to generate a general phase diagram for three variables: 
the temperature, the stretching force, and the unzipping force.  
\section{Flexible Model}\label{secIII}
We use the model from Ref. \cite{rajeev} introduced to study the 
unzipping transition of dsDNA discussed in the last section. First, we 
solve the model analytically through the generating function technique, and 
then we study numerically the behavior of finite-length systems. 
\subsection{Recursion relation and observables}\label{secIIIA}
In the absence of any force the recursion relation followed by this system is given by 
\cite{rajeev} :
\begin{equation}
\label{EQ:1}
 Z_{n+1}(x_{1},x_{2}) = \sum_{(i,j)=\pm1}Z_{n}(x_{1}+i,x_{2}+j)
                        [1-(1-y)\delta_{x_{1},x_{2}}],
\end{equation}
where $Z_{n}(x_{1},x_{2})$ is the canonical partition function of the system 
of two polymers, each of length $n$, and the spatial positions of the $n$th monomers 
of polymer $1$ and polymer $2$ are $x_1$ and $x_2$, respectively. For a given 
monomer number if $x_{1}$ becomes equal to $x_{2}$, then there is a contact. We set 
the initial condition as $Z_{0}(0,0)=y$ such that two strands start from 
the origin. The non-crossing constraint is implemented by not letting $x_1$ 
becoming greater than $x_2$ $(x_1\leq x_2)$.

We apply a constant stretching force $g_s$ at the open end point of each 
strand. The partition function of $n$-length DNA in the presence of this 
stretching force is given by
\begin{equation}
\label{EQ:2}
 {\sf Z}(g_s)=\sum_{x_{1},x_{2}} Z_{n}(x_1 , x_2) e^{g_s X},\hspace{0.1in}\text{where}\hspace{0.1in}X=x_{1}+x_{2}
\end{equation}
and the sum is over all the allowed values of $x_1$ and $x_2$.

The elastic response of the system under a stretching force can be quantified 
through the average extension $(x)$ and the elastic modulus $(\kappa)$. We 
define them in the following way
\begin{equation}
\label{EQ:3}
 x=\pd{f}{g_s}=\frac{1}{N}\pd{\ln {\sf Z}(g_s)}{g_s},\hspace{0.1in}
 \text{and}\hspace{0.1in}\kappa=\pd{x}{g_s},
\end{equation}
where $f=\beta F=-\ln Z(g_s)$ is the free energy of the system scaled by $\beta$, and 
$N$ is the length of the strands $(N\to\infty)$. Using 
Eq. \eqref{EQ:2} we can rewrite them as 
\begin{subequations}
 \begin{eqnarray}
  \label{EQ:4a}
  x &=&\frac{\langle X\rangle}{N}=\frac{\sum_{x_{1},x_{2}} Z_{N}(x_1 , x_2) e^{g_s X}
     X}{N\sum_{x_{1},x_{2}} Z_{N}(x_1 , x_2) e^{g_s X}},\\
  \label{EQ:4b}
  \kappa&=&\frac{1}{N}\left(\langle X^2\rangle - \langle X\rangle^2\right),
 \end{eqnarray}
\end{subequations}
where $\langle...\rangle$ denotes the thermal average as indicated in Eq. \eqref{EQ:4a}. 
An inspection of Eqs. \eqref{EQ:4a} and \eqref{EQ:4b} shows that $x$ is related to the average 
vectorial position of the center of mass of the end points of the two strands
of length $n$ under a stretching force $g_s$, and as expected,  $\kappa$ is 
related to the fluctuation of $x$. If $x_1$ and $x_2$ are uncorrelated, then $\kappa$ is 
the sum of the individual elastic constants. This will be the case in the unbound 
phase. According to this definition for a given force the larger the value 
of $\kappa$, the greater is the flexibility of the dsDNA. Two other important quantities 
are the average number of contacts between two strands $(n_c)$ and its 
fluctuation $(C_c)$. Two extreme values of $n_c$, 0 and 1, represent the unbound 
and the bound states, respectively. As $y$ is a temperature-like variable one can derive 
the specific heat of the system from $C_c$. We call it the specific heat for brevity. 
These are defined as 
\begin{eqnarray}
 \label{EQ:5}
 n_c = \frac{y}{N}\pd{f}{y}\hspace{0.1in}
 \text{and}\hspace{0.1in}C_c &=& y\pd{n_c}{y}.
\end{eqnarray}
We follow these definitions in the rest of this paper.

The recursion relations defined above can be solved exactly in the
infinite-chain limit (i.e., the thermodynamic limit) with the help of
generating functions.  Based on the closed-form expressions, the
physical quantities are obtained by taking appropriate derivatives as
in Eq. (\ref{EQ:3}).  The results obtained in this way are called
``analytical results.''  These recursion relations can also be
evaluated exactly, but numerically, by using the transfer matrix
technique, for finite-length chains.  The physical quantities like 
the average extension and elastic modulus are then
obtained by using Eqs. (\ref{EQ:4a}) and (\ref{EQ:4b}). To evaluate $n_c$ and 
$C_c$ numerically we need to find out the first and the second numerical 
derivatives of the partition function with respect to $y$. We evaluate 
the recursion relations for the first and the second derivatives of the 
zero-force partition function by taking the first and second derivatives of 
Eq. \eqref{EQ:1} with respect to $y$, respectively, and iterate them 
numerically to evaluate them. Then, using Eq. \eqref{EQ:5} we 
calculate $n_c$ and $C_c$ for zero applied force. For a constant 
nonzero applied force, we follow the same procedure but evaluate the 
partition function and its $y$ derivatives by using Eq. \eqref{EQ:2} 
and taking derivatives of Eq. \eqref{EQ:2} with respect to $y$. 
Such exact numerical results below are to called ``numerical results.''

\subsection{Elastic Response Under a Stretching Force}\label{secIIIB}
\stfphdiag
\subsubsection{Generating function and the free energy}
By employing the generating function technique the recursion relation 
Eq. \eqref{EQ:1} can be exactly solved. We define 
\begin{equation}
\label{EQ:6}
 G(z,x_{1},x_{2})=\sum_{n=0}^{\infty}z^n Z_{n}(x_1,x_2).
\end{equation} 
By doing this we are going to the grand-canonical ensemble from the canonical ensemble. 
By multiplying both sides of Eq. \eqref{EQ:1} by $z^n$ and then summing over $n$ we 
get two independent equations: one for nonzero unequal values of 
$x_1$ and $x_2$, and another for $x_1=x_2=0$. These are given by
\begin{subequations}
\begin{eqnarray}
\label{EQ:7a}
 \frac{1}{z}G(z,x_1,x_2)&=&\sum_{(i,j)=\pm1}G(z,x_1+i,x_2+j),\\
\label{EQ:7b}
 \frac{1}{yz}G(z,0,0)&=&\frac{1}{z}+\sum_{(i,j)=\pm1}G(z,i,j).
\end{eqnarray}
\end{subequations}
The free energy per unit length of the DNA is determined by the singularity of 
$G(z,x_1,x_2)$ closest to the origin in the complex $z$-plane.

Assuming a power-law form for $G(z,x_1,x_2)$ with respect to the relative position 
coordinate we make the ansatz
\begin{equation}
\label{EQ:8}
 G(z,x_1,x_2)=A(g_s,z)\lambda(g_s,z)^{(x_1-x_2)/2}e^{g_s(x_{1}+x_{2})},
\end{equation}
where $A$ and $\lambda$ are functions of $z$ and independent of position 
coordinates. Equation \eqref{EQ:8} generalizes the ansatz in Ref. \cite{rajeev}. 
Using the ansatz, Eq. \eqref{EQ:8}, in Eq. \eqref{EQ:7a} and 
Eq. \eqref{EQ:7b}, two unknowns, $A(g_s,z)$ and $\lambda(g_s,z)$, can easily be solved. 
Their forms are given in Appendix \ref{appnA}. 
The free energies of the different phases of the system are obtained from 
the singularities of $G(g_s,z)$. The singularity $z_b$ of $A(g_s,z)$ corresponds 
to the bound-state free 
energy and the branch point singularity $z_f$ of $\lambda(g_s,z)$ 
gives the free energy of the unbound state. They are calculated as
\begin{subequations}
\begin{eqnarray}
\label{EQ:10a}
 z_b(y,g_s)&=&\frac{y-1}{y\ \text{sech}(2g_s)}\left[\sqrt{\frac{\text{sech}^2(2g_s)}{y-1}+1}-1\right],\\
\label{EQ:10b}
 z_f(y,g_s)&=&\frac{\text{sech}^2(g_s)}{4}.
\end{eqnarray}
\end{subequations}
The difference in the force term can be understood by looking at the low-energy 
excitations. In the case of the free chains a force $g_s$ flips a bond  interchanging 
the  energies $\pm g_s$. This gives the $\text{sech}^2(g_s)$ term. In the bound state, with 
coincident end points, a bound bond gets flipped under a force $2g_s$, yielding the 
$\text{sech}^2(2g_s)$ term. From here onwards we suppress $y$ and $g_s$ as arguments 
for notational simplification and show them whenever necessary. The corresponding 
dimensionless free energies per unit length are given by 
\begin{subequations}
 \begin{eqnarray}
 \label{EQ:11}
  f_b&=&\ln z_b,\\
  f_f&=&\ln z_f.\label{eq:11b}
 \end{eqnarray}
\end{subequations}
There are two parameters in this formulation, $y$ and $g_s$.
The singularities move when these two parameters are changed.
Consider a situation where the system is in the bound state with the free 
energy given by $f_b$. Now, we can vary the parameters in such a way 
that the unbound-state singularity $z_f$ crosses $z_b$ and becomes 
closest to the origin. In this situation the free energy of the system 
becomes $f_f$. 
The crossing of the singularities defines the transition point from bound 
to unbound by the force at
\begin{equation}
 \label{EQ:12}
 g_{sc}=\frac{1}{2} \cosh ^{-1}\left(\frac{2-y}{2 (y-1)}\right).
\end{equation}
The phase diagram in the $y$-$g_{sc}$ plane is shown in Fig. \ref{fig2}. 
The phase boundary has the following limiting forms:
\begin{subequations}
\begin{eqnarray}
\label{ymodel1}
 g_{sc}&\sim&\sqrt{y_c-y}\hspace{0.2in}\text{for}\hspace{0.2in}y
 \rightarrow y_c-,\hspace{0.2in}\text{and}\\
 \label{ymodel2}
 g_{sc}&\sim&-\frac{\ln(y-1)}{\ln y}\hspace{0.2in}\text{for}
 \hspace{0.2in}y\rightarrow 1+.
\end{eqnarray}
\end{subequations}
\subsubsection{Results and discussions}
\paragraph{Long-chain limit.}
\stfnc
When the two 
strands are in the unbound state, they come closer and form contacts in the 
influence of the stretching force. In this way an unbound state becomes a bound 
state above the critical stretching force. Figure \ref{fig1} shows how $n_c$, 
calculated using Eq. \eqref{EQ:5}, 
becomes non-zero before saturating at 1 as the stretching force crosses a 
critical value for a $y<y_c$. This shows that the transition is 
continuous. It is already 
known that at the point $(4/3,0)$ in the phase diagram the system goes through 
a second-order phase transition. Beyond this critical point the system always 
remains in the bound state, thus excluding any other possibilities of phase 
transition. The only effect of the stretching force there is to influence  
the bubble statistics. The asymptotic behavior of $n_c$ is given by  
\begin{equation}
n_c\approx \begin{cases}
\frac{9}{2}(g-g_{sc})\sqrt{y-y_c}, & {\rm for \ }  \mycom{y\to y_c +,}{g_s\to g_{sc}+,}\\
\frac{3}{2} \ g_s^2, & {\rm for \ } y=y_c,  g_s\to 0,\\
n_0 + \frac{g_s^2}{\sqrt{y(y-1)}}, & {\rm for \ } y>y_c, g_s\to 0,\\
\frac{27}{8} \left(y-y_c\right), & {\rm for \ } y\to y_c+, g_s= 0,
\end{cases}
\end{equation}
where 
\begin{equation}
n_0=n_c(y>y_c,g_s=0)=\frac{y-2+\sqrt{y(y-1)}}{2(y-1)}.
\end{equation}
That this is a second-order phase transition can be corroborated by examining the 
average extension of the center of mass due to the application of the 
stretching force and the elastic modulus. They are calculated using the 
definitions of Eq. \eqref{EQ:3}. 
\stfext
Figure \ref{fig3} shows how the average extension changes continuously as the 
system crosses the critical stretching force. 
For a zero force $x$ is 0, consistent with the Gaussian chain behavior, 
while the fully stretched state under a large force has $x=2$.
\ananumcoma
\ananumcomb
The slope discontinuity at the transition point $g_{sc}$ of Eq. \eqref{EQ:12} 
gives rise to a jump in the elastic constant as shown in Fig. \ref{fig4a}. 
To be noted here is that there is no pretransitional signature on either 
side of the transition. 
\figyzerog
However, for a finite-size system 
the scenario is different. All other curves in Fig. \ref{fig4a} except 
the analytical curve are for different finite sizes of the system. In the 
unbound phase each strand has the equation of state $x=\tanh(g_s)$ so that 
the total $x=2\tanh(g_s)$. This is the $g_s < g_{sc}$ branch. The corresponding 
entropic elastic constant is $\kappa=2\ \text{sech}^2(g_s)$. The completely 
bound state, in the absence of any bubbles, should have a similar equation of 
state, with an elastic constant of purely entropic origin given by 
$\kappa=4\ \text{sech}^2(2g_s)$. But the bubbles 
give an extra contribution. The exact form of the elastic constant can be 
determined for a few special cases. The $y$ dependence of the zero force 
$\kappa$ is given by (see Fig. \ref{figy1})
\begin{equation}\label{eq:1k}
\kappa(g_s=0)=\begin{cases}
2, & \text{for}\quad y<y_c,\\
4\sqrt{\frac{y-1}{y}}, & \text{for}\quad y>y_c.
\end{cases}
\end{equation}
The elastic constant as a function of force at the melting point $y=y_c$ is 
\begin{equation}\label{eq:2c}
\kappa(y=y_c)=\frac{64 w \left(w+1\right)}{\left(w^2+14 w+1\right)^{3/2}},
\ \text{with}\quad w=e^{4g_s}.
\end{equation}
The behavior of $\kappa$ for $y=y_c$ and $y>y_c$ is shown in Fig. \ref{figy2}. 
\figgsdy
\paragraph{Finite-length DNA.}\label{sec:finite-length-dna}
The contribution of the bubbles becomes significant in finite-length DNA as 
shown in Fig. \ref{fig4b}. The finite-size effects become significant when 
the length is comparable to or shorter than the length of the bubble fluctuations. 
The elastic constant for a finite-length DNA is necessarily continuous, 
devoid of any singularity, but it should evolve into a discontinuous function 
as the length is increased. This indicates that shorter chains will show 
a larger deviation from the thermodynamic limit over a range of forces. A finite-size 
scaling form is 
\begin{equation}
\label{scaling}
\kappa={\sf f}((g_s-g_{sc})N^{1/\nu}), 
\end{equation}
so that $\kappa={\sf f}(0)$ at $g_s=g_{sc}$ for all finite $N$. Therefore all the 
finite-size curves pass through a common point as shown in Fig. 
\ref{fig4b}, which is the critical point. By identifying the common points 
for other $y$ values we can now determine the phase diagram numerically. This 
is shown in Fig. \ref{fig2}. The consistency between this numerical method 
of finding the critical points with the analytical results helps us when the 
model under consideration is not solvable analytically. 
All the points in this phase boundary including 
the thermal melting point $(y=4/3,g_{sc}=0)$ are second-order critical points. The behavior 
of $\kappa$ shows that the system is most flexible when it is in the unbound state 
and under no external force, as it has the highest value of $\kappa$. 
\subsection{Role of the bubbles}
To highlight the importance of the bubbles we compare our results 
with the Y-model which is similar to the flexible model except that bubble formation 
is not allowed there \cite{marenduzzo}. The bound state of this model is the same as the 
completely bound state of the flexible model and it has a zero-force melting point 
(a first-order transition) at $y_c=$2. In the presence of $g_s$ the corresponding 
singularities and elastic constants are given by 
\begin{subequations}
 \begin{eqnarray}
 \label{ymod1}
  z_b&=&\frac{1}{2y\cosh2g_s},\quad\kappa_b=4\ \text{sech}^2(2g_s),\hspace{0.1cm}\text{and}\\
  z_f&=&\frac{1}{2\cosh^2g_s},\quad\kappa_f=2\ \text{sech}^2(g_s),
 \end{eqnarray}
\end{subequations}
where $\kappa_b$ and $\kappa_f$ are the bound-state and the unbound-state elastic constants, 
respectively. We obtain the phase boundary by equating $z_b$ with $z_f$ as
\begin{equation}
 g_{sc}=\frac{1}{2}\cosh^{-1}\left(\frac{1}{y-1}\right).
\end{equation}
The phase boundary has similar asymptotics for $y\to y_{c}(=2)$ and $y\to1$ 
as in Eqs. \eqref{ymodel1} and \eqref{ymodel2}. 
In Fig. \ref{fig4a} we compare $\kappa_b$ with the flexible model results.
It shows that the flexibility of the bound state of the flexible model is 
mostly due to the bubbles.
\subsection{Elastic response in the presence of an unzipping force}\label{secIIIC}
It is well known that this model undergoes an unzipping transition under the 
influence of an unzipping force in the absence of a stretching force. This 
unzipping transition is known to be a first-order phase transition. 
The unzipped state consists of two completely separated independent 
single strands. When DNA is in the double-stranded form the unzipping 
force tries to unzip it into two single strands. On the other hand, when 
DNA is in the unzipped state the stretching force tries to make them 
bound. Now, if 
we apply both the forces simultaneously we expect a competition between 
the opposing effects. In this section we study this problem, again, 
analytically for the infinite system and numerically for finite systems.
We use the same definitions of quantities as in Eqs. \eqref{EQ:4a}, 
\eqref{EQ:4b}, and \eqref{EQ:5}.

Let us apply a spatially independent unzipping force $g_u$ at the rear 
end of the DNA, i.e., the forces act exactly in opposite directions. In the presence of a 
stretching force $g_s$ the generating function is given by 
$G(g_s,z)=A(g_s,z)\lambda(g_s,z)^{(x_1-x_2)/2}e^{g_s(x_1+x_2)}$, where $A(g_s,z)$ and 
$\lambda(g_s,z)$ are given by Eq. \eqref{EQ:9a} and Eq. \eqref{EQ:9b}. The generating 
function in the presence of both forces is given by 
\begin{equation}
 \label{EQ:13}
 {\cal G}(g_s,g_u,z)=\sum_{x_1,x_2}G(g_s,z)e^{g_u (x_1-x_2)}.
\end{equation}
So, the bound-state singularity remains the same as $z_b$, Eq.
(\ref{EQ:10a}), consistent with the hypothesis of non-penetration of
forces in the bound state \cite{poulomi}, but the unbound-state
singularity is now given by the solution of the equation $e^{2
  g_u}=\lambda(g_s,z)$. This equation is easily obtained by performing
the summation over $x_1$ and $x_2$ in Eq. \eqref{EQ:13}. Solving this
equation for $z$ we find that the unbound-state singularity $z_{fu}$ is
given by
\begin{equation}
 \label{EQ:14}
 z_{fu} = \frac{1}{2\left[\cosh(2 g_s)+\cosh(2 g_u)\right]},
\end{equation}
which corresponds to the partition function of the two chains under forces 
$g_s+g_u$ and $g_s-g_u$, namely, $4\cosh(g_s+g_u)\cosh(g_s-g_u)$. For $g_u=0$, 
the corresponding singularity matches Eq. \eqref{EQ:10b}. The transition, 
as before, is given by the crossing of the singularities at  
\begin{equation}
 \label{EQ:15}
 g_{uc}=\frac{1}{2} \cosh ^{-1}\left(\frac{1}{2 z_b}
         -\cosh \left(2 g_{\text{s}}\right)\right).
\end{equation}
This expression reduces to the known unzipping line \cite{marenduzzo} for $g_s=0$ and 
Eq. \eqref{EQ:12} for $g_u=0$.
\threedphdiag
\subsubsection{Complete phase diagram with unzipping force}
After the introduction of an unzipping force we now have three control
parameters. By changing any one of them while keeping the other two fixed, one can 
induce a phase transition in the system. The transition points 
are distributed on a surface in the $y$-$g_{s}$-$g_{u}$ 
space given by Eq. \eqref{EQ:15}. In Fig. \ref{fig7} we plot this function. 
The critical curve for $g_{uc}=0$ in the surface is a second-order line. 
Around $g_s=0$, $g_{uc}(g_s)=g_{uc}(0)+a g_{s}^2+...$, so 
that the unzipping line for $g_s=0$ lies along the locus of the local minima 
on the surface. The first-order surface ends on the $g_u=0$ plane in a 
critical line that contains the usual melting point at $y_c(g_s=g_u=0)$. 
Except for the critical line all the other possible lines on the surface are first-order 
lines. To show that there is indeed a first-order transition we plot 
$n_c$ as a function of $g_s$ in Fig. \ref{fig1} with $g_u >0$ and $y$ 
kept fixed. 
\unzanaext
\unzanelc
\subsubsection{Elastic constant}\label{sec:elastic-constant}
We plot $x$ vs $g_s$ in Fig. \ref{fig6a} and $\kappa$ vs $g_s$ in Fig. \ref{fig6b}, 
keeping $g_u$ and $y$ at fixed values. The dotted magenta curves are the analytical ones 
for an infinite system length, obtained by using Eqs. (\ref{EQ:3}),
(\ref{EQ:10a}), and (\ref{EQ:14}), while all the other curves are for finite system sizes which 
gradually match the analytical curve as $N$ becomes larger. $x$ shows a finite 
discontinuity at a critical $g_{sc}=1.18$. The analytical curve for $\kappa$ has a 
$\delta$-function peak at $g_{sc}$, which is not shown in Fig. \ref{fig6b}. The 
uniform increase in the peak height with increasing system size 
in $\kappa$ at the critical point is the signature of the delta peak. 
Below we list various useful limiting values of $x$ and $\kappa$. 
\begin{enumerate}
\item $\text{For}\quad g_s\to0,g_u>g_{uc}(y)$, 
\begin{subequations}
 \begin{eqnarray}
  x &\approx& 2\ \text{sech}^2\left(g_u\right)g_s,\\ 
\kappa &\approx& 2\ \text{sech}^2\left(g_u\right)+2 \frac{\cosh(g_u)-2}{\cosh^4(g_u)}g_s^2.
 \end{eqnarray}
\end{subequations} 
\item $\text{For}\quad g_s\to0,g_u<g_{uc}(y),$
\begin{subequations}
 \begin{eqnarray}
  x&\approx&4 \sqrt{\frac{y-1}{y}} g_s,\\
\kappa&\approx&4 \sqrt{\frac{y-1}{y}}+\frac{8 (3-2 y) \sqrt{y-1}}{y^{3/2}}g_s^2.
 \end{eqnarray}
\end{subequations} 
\item $\text{For}\quad g_s\to g_{sc}-,y=1.2,g_u=0.4$, 
\begin{subequations}
 \begin{eqnarray}
  x &\approx& 1.57107 + 0.73049 (g_s-g_{sc}),\\ 
\kappa &\approx& 0.73049 - 1.03646 (g_s-g_{sc}).
 \end{eqnarray}
\end{subequations} 
\item $\text{For}\quad g_s\to g_{sc}+,y=1.2,g_u=0.4$, 
\begin{subequations}
 \begin{eqnarray}
  x &\approx& 1.81211 + 0.66067 (g_s-g_{sc}),\\ 
\kappa &\approx& 0.66067 - 2.01486 (g_s-g_{sc}).
 \end{eqnarray}
\end{subequations}
\end{enumerate}
For an infinite system the transition occurs suddenly 
at a single point. On the other hand, for a finite-size system the 
effect of the transition remains relevant for a domain of $g_s$ 
values containing $g_{sc}$ beyond the scaling regime. 
\section{Rigid Model}\label{secIV}
Here we customize the previous model to incorporate 
explicit weights for bubble formation. Doing that in the transfer 
matrix format is a bit involved. To identify a bubble we need to 
ensure that an unbound region is attached between two bound segments. 
A bound segment is defined as a DNA patch where every base pair is 
in the bound state and the minimum length it can have is 2. We 
implement this by applying the constraint that a bound base pair can 
form only if another bound base-pair precedes it. So, for every 
step in the generation of the polymers we need to keep track of 
the previous step. We introduce a built-in rigidity to the dsDNA 
by instituting a bias against the bending towards the right of bound 
segments. For computational simplicity here we completely 
switch off the rightward option.
By doing this we are introducing a bias in the propagation of 
DNA in favor of one direction. Other than the usual contact weight 
$y$ we introduce another Boltzmann weight $v$ if a bound segment 
opens to form two single strands or two single strands recombine 
to form a bound segment. The recursion relation which obeys these 
rules is given by 
\begin{widetext}
\begin{equation}
\label{EQ:16}
z_n(x_1,x_2)=\\
 \begin{cases}
 \text{\color{blue}y} [\text{\color{red}v} z_{n-1}(i,l)
 +\text{\color{red}v} z_{n-1}(j,k)+z_{n-1}(j,l)], 
 & \text{if}\hspace{0.1in} x_1=x_2\&
   n>0 \\
 \text{\color{red}v} z_{n-1}(i,l)+z_{n-1}(i,k)
 +z_{n-1}(j,k)+z_{n-1}(j,l), &
   \text{if}\hspace{0.1in}x_1-x_2=2\& n=1 \\
 \text{\color{red}v} z_{n-1}(j,k)+z_{n-1}(i,k)
 +z_{n-1}(i,l)+z_{n-1}(j,l), &
   \text{if}\hspace{0.1in}x_1-x_2=-2\& n=1 \\
 \text{\color{red}v} \text{\color{blue}y} z_{\text{\color{red}n-2}}(i+1,l+1)
 +z_{n-1}(i,k)+z_{n-1}(j,k)+z_{n-1}(j,l), &
   \text{if}\hspace{0.1in}x_1-x_2=2\& n\geq 2 \\
 \text{\color{red}v} \text{\color{blue}y} z_{\text{\color{red}n-2}}(j
 +1,k+1)+z_{n-1}(i,l)+z_{n-1}(i,k)+z_{n-1}(j,l), &
   \text{if}\hspace{0.1in}x_1-x_2=-2\& n\geq 2 \\
 z_{n-1}(i,k)+z_{n-1}(i,l)+z_{n-1}(j
 ,k)+z_{n-1}(j,l), &
   \text{if}\hspace{0.1in}\left| x_1-x_2\right| >2\& n>0
\end{cases},
\end{equation}
\end{widetext}
with $i=x_1-1$, $j=x_1+1$, $k=x_2-1$ and $l=x_2+1$. 
The first two steps fix the initial configurations. To fix the configuration 
at the $n$th step we need to keep the information on not only the $(n-1)$th step 
but also the $(n-2)$th step. 
We evaluate the recursion relation, Eq. \eqref{EQ:16}, exactly for finite system 
sizes by iterating it numerically. Once $Z(x_1,x_2)$ is known, the 
force-dependent partition function can be obtained with the help of Eq. \eqref{EQ:2}, 
and the corresponding elastic constant from Eq. \eqref{EQ:3}. Only the stretching 
force $g_s$ is considered here. Using the bubble weight 
$v$ we can now count the average number of bubbles $(n_b)$ and calculate the 
average bubble length $(l_b)$. They are formally given by the formulas 
\begin{equation}
 \label{EQ:17}
 n_b = \frac{v^2}{N}\pd{f}{v^2},\hspace{0.1cm}C_b = v^2\pd{n_b}{v^2}\hspace{0.1cm},
 \text{and}\hspace{0.1cm}l_b=\frac{(1-n_c)}{n_b},
\end{equation}
where $C_b$ describes the fluctuations in $n_b$. To evaluate them numerically, first 
we take the first and second derivatives of Eq. \eqref{EQ:16} with respect to $v$ and 
iterate them numerically by setting $v=1$ to obtain the first and the second derivatives 
of the zero-force partition function respectively. Once the derivatives of the partition 
function are known, $n_b$, $C_b$, and $l_b$ can be determined by using Eq. \eqref{EQ:17} for 
a zero applied force. To find out the corresponding quantities for nonzero constant 
forces and to find out $n_c$ and $C_c$, we follow the method described in Section \ref{secIIIA}.

\subsection{Thermal Melting : $g_s=0$}\label{secIVA}
First, let us show that this model goes through a binding-unbinding transition 
as $y$ is varied in the absence of any external forces. For the analysis in 
this section we set $v=1$. Here we use the exact numerical transfer matrix 
method for finite-size systems.

In Fig. \ref{fig8} we show how the average number of contacts varies with 
$y$. As the system size is increased one part of the curve gradually touches the 
$y$ axis. And it is also evident that $n_c$ will saturate at $n_c=1$ for 
appropriately high $y$ values. These indicate a binding-unbinding transition. 
To find out the order of the transition and the corresponding critical 
value of $y$, $y_c$, we plot $C_c$ vs $y$ in Fig. \ref{fig9}. $C_c$ 
obeys a finite size scaling relation similar to Eq. \eqref{scaling} which 
indicates a finite discontinuity. At $y=1.18$ all the curves 
pass through a common point, implying $y_c=1.18$. The finite discontinuity at $y_c$ 
establishes that this is a second-order phase transition. 
\rcnc
\rcsp
Note that we have not imposed the 
non-crossing constraint here. The restriction imposed on the bound state 
is sufficient to induce a bound-unbound transition. In one spatial 
dimension the entropy of a system dominates over binding energy, which 
implies that there is no ordering here. Imposition of special restrictions 
to limit the entropy may result in an energy dominated ordered state. The 
non-crossing constraint in the previous model did exactly that by 
decreasing the total number of configurations. The restriction imposed 
here on the degrees of freedom of the bound segment plays a similar 
role, decreasing the total number of configurations of the DNA. We elaborate 
on this in Appendix \ref{appnB}.
\subsection{Elastic response : $g_s\neq0$}\label{secIVB}
Let us now discuss the elastic properties of this system. As discussed 
earlier the inherent asymmetry in this model favors 
extension of DNA in one direction and opposes it in the other direction. So 
under the influence of a spatially independent stretching force the DNA is 
more flexible in one direction compared to the other direction.
\rcext
\rcelc
As the 
system is no longer symmetric under $g_s\leftrightarrow -g_s$ we need to 
focus attention on the negative values of $g_s$ too. Figure \ref{fig10} shows that 
$x$ varies continuously with $g_s$, reaching $\pm 2$ for a large positive or negative 
$g_s$. In Fig. \ref{fig11} we plot $\kappa$ vs $g_s$ keeping $y$ fixed. Every 
curve shows a peak around $g_s=0.02$ which increases in height as $N$ increases. 
The maximum of $\kappa$, $\kappa_{\text{max}}$, goes to a finite value in the $N\to\infty$ 
limit as shown in the inset in Fig. \ref{fig11}. The inset in Fig. \ref{fig12} 
shows how $n_c$ changes as we increase $g_s$ for $y=1.20$. This 
indicates a continuous binding-unbinding phase transition. Figure \ref{fig12} shows 
that $C_c$ has a finite jump at $g_{sc}=0.02$ which can be identified as the 
common point in the peak region through which every finite-size curve passes.
For $g_s<g_{sc}$, $\kappa$ shows anomalous behavior, as close to the transition point 
it can reach values which are much greater than the entropic elastic modulus 
of the unbound state given by $2\ \text{sech}^2(g_s)$, shown in Fig. \ref{fig11} 
as a dashed black line.
\rcfnc
\rcfphd
By collecting 
similar common points for different $y$ values we draw the numerical phase 
diagram of the system, which is shown in Fig. \ref{fig14}. Noticeable 
there is that the stretching force actually unbinds the bound state for $g_s>g_{sc}$. The 
reason for this is the following. Due to the bias the bound-state formation 
on the positive $x$ axis is unfavorable and the DNA prefers to go in the 
negative $x$ direction. As $g_s$ is increased it wins over the bias eventually 
and pulls the DNA towards the positive $x$ direction. Because the bound state is 
forbidden in that direction the bound DNA unzips as a result.   
\subsection{Role of the bubbles}\label{secIVC}
The flexibility of the bound state comes solely from the bubbles, as the 
bound segments are absolutely rigid in this model.
\rcfnblb
\rcfnbcb
\xnr
\kcr
Figure \ref{fig15} shows that $n_b$ becomes very 
small while $l_b$ increases to almost equal to $N$ as we approach the transition 
point. Here $l_b\approx (N\ \text{or}\ 0)$ means the DNA is in the unzipped state. The peaks in $n_b$ 
curves indicate a large number of bubbles but at the same time of very 
small average length. From these two observations we can now say that as we 
approach the transition point many small bubbles coalesce to form large 
bubbles with decreasing numbers, and eventually $n_b$ becomes 0 when the two 
strands get completely separated. The fluctuation in $n_b$, $C_b$, also 
becomes large around the critical point as shown in Fig. \ref{fig16}. 
Earlier we have shown that $\kappa$ in this model behaves anomalously and the 
anomalous behavior occurs in the same region where $l_b$ and $C_b$ are the largest. 
In our model, $\kappa$ is the fluctuation of extension $x$ by 
definition and it depends on $n_b$. For example, near the transition point 
$n_b$ is very small and $x$ is also very small, although $l_b$ is large. This is 
because in the $(1+1)$ dimension $x$ for a single strand is 0 due to its 
Gaussian nature. But for a bound DNA, $x\propto\sqrt{n_b}$ as shown in Fig. 
\ref{fig17}. It is then 
expected that $\kappa$ will be determined by the fluctuations in $n_b$, $C_b$. 
In Fig. \ref{fig18} we plot $\frac{1}{\kappa}\sqrt{C_b}$ vs $y$ for 
different system sizes in the absence of any external force. For the bound region 
$(y>1.18)$ the curves collapse into a single master curve which 
is almost $y$ independent, inferring that 
\begin{equation}
\kappa\approx7.7\sqrt{C_b}.
\end{equation}
We therefore conclude that $C_b$ is the important factor in determining the elastic 
behavior of the system.
\section{Discussion And Summary}\label{secV}
There are a few points which we feel need to be clarified in more 
detail. 
(a) As the free ends of the  
two strands are stretched more and more with increasing forces in 
the same direction, they are bound to come closer due to their equal 
lengths and the starting ends' being attached to each other.  
This coming closer together increases the possibility of the strands' 
forming a bound base pair by gaining energy. 
(b) The transition in the presence of 
the unzipping force is definitely an unzipping transition. This is 
true even if $g_u$ is very small. 
(c) $\kappa$ in the flexible model is sensitive to the changes in $g_u$, 
$g_s$, and $y$. Elasticity is entropic in nature, which emerges from 
collective behavior. The rigid model, on the other hand, has its own intrinsic 
elasticity. This is reflected in the anomalous behavior of $\kappa$ for 
this model. 
(d) The single-molecule DNA experimental setup in which stretching force is 
achieved by placing the DNA in a directional flow can be a testing platform 
of our models. 
(e) In the nanopore sequencing technique, dsDNA is unzipped and a single 
strand is passed through a nanopore \cite{nanopore}. Other than that, during 
bacterial conjugation or infection of a cell by a virus the DNA assumes 
similar geometry. Our study may be relevant in these cases. 
(f) Single molecule DNA unzipping experiments are normally done at 
room temperature. In Eq. \eqref{EQ:15} we provided a phase boundary 
which depends not only on the temperature and the unzipping force, but also 
on the stretching force. Its remains a challenge to generate this phase 
boundary experimentally with temperature as a variable in single-molecule 
experiments.

We summarize the basic results on rigidity as defined by Eq. \eqref{EQ:3} 
for the two models. 

For the entropic rigidity as seen in the flexible-chain model of DNA, we 
obtained the following {\it exact} results.
(1) For $g_s=g_u=0$,
\begin{equation}
\kappa(g_s=0)=\begin{cases}
2, & \text{for}\quad y<y_c,\\
4\sqrt{\frac{y-1}{y}}, & \text{for}\quad y>y_c,\nonumber
\end{cases}
\end{equation}
(2) For $y=y_c$, i.e. at the critical point,
\begin{equation}
\kappa(y=y_c)=\frac{64 w \left(w+1\right)}{\left(w^2+14 w+1\right)^{3/2}},
\ \text{with}\quad w=e^{4g_s}.\nonumber
\end{equation}
In the presence of the two opposing forces, $g_s$ as the stretching force 
and $g_u$ as the unzipping force, the transition surface in the 
$y$-$g_s$-$g_u$ space is given by Eq. \eqref{EQ:15}.

For the model with intrinsic rigidity, the main result we obtained is 
\begin{equation}
\kappa\approx7.7\sqrt{C_b}.\nonumber
\end{equation}
\section{Conclusion}\label{secVI}
We have studied the effect of melting of DNA on its elasticity using 
$(1+1)$ dimensional models by employing exact numerical and analytical 
methods. Under a stretching force DNA goes through a second-order 
binding-unbinding phase transition. The dependence of DNA flexibility 
on the stretching force, the unzipping force, and the temperature has also been 
discussed. In the presence of both forces the system goes through 
a first-order unzipping transition. The complete phase diagram in the 
$y$-$g_s$-$g_u$ space is obtained. The average bubble length and the average 
bubble number for our model for different parameter values are also studied. 
We have shown that the DNA flexibility is related to the 
bubble number fluctuations. For zero external forces, the extension of the 
DNA is temperature independent and varies with the square root of the bubble numbers 
proportionally, while the elastic modulus is also proportional to the square 
root of the bubble number fluctuation. Though the binding-unbinding transition is 
very sharp for an infinite-length system, the transition point can 
influence the elastic behavior of DNA for a broad region of parameter 
values when the system length is finite. Consequently, the elastic response 
of short-length DNA, as used extensively in experiments, has to be widely 
different from that of long-chain DNA. Furthermore, though DNA is a very long molecule, 
it can melt locally, depending on the environment it is in. Thus our study 
will help us to understand the importance of these locally melted regions of 
shorter lengths in determining the elastic properties of the DNA as a whole. 
\appendix
\section{$A(g_s,z)$ AND $\lambda(g_s,z)$}\label{appnA}
Forms of $A(g_s,z)$ and $\lambda(g_s,z)$ needed for $z_b$ and $z_f$ (Eqs. \eqref{EQ:10a} 
\eqref{EQ:10b}) are given by 
\begin{widetext}
\begin{subequations}
\begin{eqnarray}
\label{EQ:9a}
 A(g_s,z)&=&\frac{-1/(2z)}{\cosh(2g_s)- 
       \sqrt{\left(\cosh(2g_s)-\frac{1}{2z}\right)^2-1}+\frac{(y-2)}{2yz}},\\
\label{EQ:9b}
\lambda(g_s,z)&=&\sqrt{\left(\cosh(2g_s)-\frac{1}{2z}\right)^2-1}+\frac{1}{2z}-\cosh(2g_s).
\end{eqnarray}
\end{subequations}
\end{widetext}
\section{BIAS-INDUCED MELTING}\label{appnB}
\biasednc
\biasedsp
In the rigid model we have completely switched off one degree of freedom 
for the bound segments. We can do the entropy limiting job in a more general 
way by introducing a control parameter $b$ instead, such that for $b=0$ we 
block a degree of freedom of the bound state completely, for $b=1$ we get back the 
good old free Gaussian chain, and for the intermediate values of $b$ we obtain a 
partially biased scenario. The recursion relation followed by the system is now 
given by 
\begin{eqnarray}
\label{EQ:x1}
 Z_{n+1}(x_{1},x_{1}) &=& y[Z_{n}(x_{1}+1,x_{1}+1) + Z_{n}(x_{1}+1,x_{1}-1)\nonumber\\
                      &&+ Z_{n}(x_{1}-1,x_{1}+1) + b Z_{n}(x_{1}-1,x_{1}-1)],\nonumber\\
 Z_{n+1}(x_{1},x_{2}) &=& [Z_{n}(x_{1}+1,x_{2}+1) + Z_{n}(x_{1}+1,x_{2}-1)\nonumber\\
                      &&+ Z_{n}(x_{1}-1,x_{2}+1) + Z_{n}(x_{1}-1,x_{2}-1)]\nonumber\\
                      &&\text{for}\hspace{0.2cm}x_1\neq x_2,
\end{eqnarray}
where $Z_{n}(x_{1},x_{2})$ is the partition function for the system length $n$. We set 
the initial condition as $Z_{0}(0,0)=y$ such that two strands start from 
the origin. The microscopic parameter $b$ is a Boltzmann weight which controls 
the possibility of the two polymers' both going to the right-hand side while 
being in the bound state. All the notations and definitions of the observables 
which are used in this model are the same as for the flexible model. By $z$-transforming 
Eq. \eqref{EQ:x1} using the initial condition we get two independent equations 
for $x_1\neq x_2 \neq 0$ and $x_1=x_2=0$: 
\begin{subequations}
\begin{eqnarray}
\label{EQ:x2a}
 \frac{1}{z}G(z,x_1,x_2)&=&G(z,x_1+1,x_2+1)+G(z,x_1-1,x_2+1)\nonumber\\
                          &&+G(z,x_1+1,x_2-1)\nonumber\\
                          &&+ G(z,x_1-1,x_2-1),\\
\label{EQ:x2b}
 \frac{1}{yz}G(z,0,0)&=&\frac{1}{z}+G(z,1,1)+G(z,-1,1)\nonumber\\
                          &&+G(z,1,-1)+ b G(z,-1,-1).
\end{eqnarray}
\end{subequations}
To solve these independent equations we make an ansatz for the generating 
function, $G(z,x_1,x_2)=A(z)\lambda(z)^{\text{Abs}[(x_1-x_2)/2]}$. Substituting 
this ansatz in Eq. \eqref{EQ:x2a} and Eq. \eqref{EQ:x2b} and solving for $A$ and 
$\lambda$ we get 
\begin{subequations}
 \begin{eqnarray}
  \label{EQ:x3}
  A&=&\frac{1}{-b z+\frac{1}{y}+z+\sqrt{1-4 z}-1},\\
  \lambda&=&-\frac{2 z+\sqrt{1-4 z}-1}{2 z}.
 \end{eqnarray}
\end{subequations}
The bound-state and unbound-state singularities are given by
\begin{subequations}
 \begin{eqnarray}
  \label{EQ:x4}
  z_b&=&\frac{b-1-y(b+1)+\sqrt{y} \sqrt{(b+1)^2 y-4 b+4}}{(b-1)^2 y},\\
  z_f&=&\frac{1}{4}.
 \end{eqnarray}
\end{subequations}
From these singularities all the other relevant quantities can be derived. 
Figure \ref{figx1} shows how $n_c$ varies with $y$. At a high enough $y$, $n_c$ 
saturates to its maximum value, 1. In Fig. \ref{figx2} we plot $C_c$ vs 
$y$. In both Fig. \ref{figx1} and Fig. \ref{figx2} we also show 
the corresponding numerical data for different finite system sizes. $C_c$ 
obeys a finite-size scaling form $C_c=g((y-y_c)N^{1/\delta})$ such that 
at $y=y_c=1.142$, $C_c(y_c)=g(0)$. The numerical data are consistent 
with the analytical results. From these two figures we conclude that 
the system is going through the usual second-order binding-unbinding phase 
transition. We obtain the critical point of this transition analytically 
by matching $z_b$ with $z_f$. 
The critical value of $y$ is now $b$ dependent and varies with $b$ as 
$y_c=4/(3+b)$. For $b=1$ we have the same recursion relation as that of a 
system with two Gaussian chains which can freely cross each other and in which there 
is no phase transition. In our case also we get $y_c=1$, which means that there is no 
phase transition at finite temperature. Another interesting limit is when 
$b=0$, $y_c=4/3$, the 
critical point for two Gaussian chains with noncrossing constraint, although 
the recursion relations for these two cases are not the same. 
Moreover, the $b$ dependence of $y_c$ in our model adds extra flexibility 
in that we can now tune the critical point by tuning $b$ for a wide range of 
values.

From the recursion relation it is clear that the parameter $b$ just modulates 
one of the four possible contributions to the partition function of the $n$th 
generation from the $(n-1)$th generation. $b$ may be associated with any of 
the four possible contributions. The case when $b$ is attached to the 
$z(x_1-1,x_2+1)$ term is of special interest because the $b\rightarrow0$ limit 
is exactly the flexible noncrossing case now. This current case also is 
exactly solvable through the generating function technique and gives  
exactly the same $b$-dependent critical melting point, $y_c=4/(3+b)$. So, we 
can now actually modulate the noncrossing constraint through $b$ and the 
corresponding critical point as well. 
%
%
%

%
%

\begin{thebibliography}{99}
\bibitem{sdna} For the inextensibility of the chains there can not be an S-DNA phase in our model as 
described in \cite{sdna0,sdna1,sdna2}.
\end{thebibliography}
\end{document}